\documentclass[12pt]{article}
\usepackage[cp1251]{inputenc}
\usepackage[english,russian,ukrainian]{babel}

\begin{document}

\vskip 0.5cm
\begin{center}
\begin{Large}
\textbf{General form of the covariant field equations of arbitrary spin and the relativistic canonical quantum mechanics}
\end{Large}
\end{center}

\vskip 0.5cm
\begin{center}
\textbf{V.M. Simulik}
\end{center}

\begin{center}
\textit{Institute of Electron Physics of the National Academy of Sciences of Ukraine, 21 Universitetska Str., 88000 Uzhgorod, Ukraine,}

e-mail: vsimulik@gmail.com
\end{center}

\vskip 1.cm

\begin{center}
\textbf{Abstract}
\end{center}

\textit{The investigation of arXiv 1409.2766v2 [quant-ph] has been continued by the general form of the numerous equations with partial values of arbitrary spin, which were considered in above mentioned preprint. The general forms of quantum-mechanical and covariant equations for arbitrary spin together with the general description of the arbitrary spin field formalism are presented. The corresponding relativistic quantum mechanics of arbitrary spin is given as the system of axioms. Previously ignored partial example of the spin s=(0,0) particle-antiparticle doublet is considered. The partial example of spin s=(3/2,3/2) particle-antiparticle doublet is highlighted. The new 64 dimensional Clifford--Dirac algebra over the field of real numbers is suggested. The general operator, which transformed the relativistic canonical quantum mechanics of arbitrary spin into the locally covariant field theory, has been introduced. Moreover, the study of the place of the results given in arXiv 1409.2766v2 [quant-ph] among the results of other authors is started. The review of the different investigations in the area of relativistic canonical quantum mechanics is given and the brief analysis of the existing approaches to the covariant field theory of arbitrary spin is initiated. The consideration of some important details of arXiv 1409.2766v2 [quant-ph] is improved.}



\vskip 0.5cm

\begin{center}
\textbf{1. Introduction}
\end{center}

\vskip 0.3cm

Recently in the arXiv preprint [1] the interesting results in the area of relativistic quantum mechanics and quantum field theory have been presented. Briefly the list of these results is as follows. The new relativistic equations of motion for the particles with spin s=1, s=3/2, s=2 and nonzero mass have been introduced. The description of the relativistic canonical quantum mechanics (RCQM) of the arbitrary mass and spin has been given. The link between the RCQM of the arbitrary spin and the covariant local field theory has been found. The manifestly covariant field equations that follow from the quantum mechanical equations, have been considered. The covariant local field theory equations for spin s=(1,1) particle-antiparticle doublet, spin s=(1,0,1,0) particle-antiparticle multiplet, spin s=(3/2,3/2) particle-antiparticle doublet, spin s=(2,2) particle-antiparticle doublet, spin s=(2,0,2,0) particle-antiparticle multiplet  and spin s=(2,1,2,1) particle-antiparticle multiplet have been introduced. The Maxwell-like equations for the boson with spin s=1 and mass $m>0$ have been introduced as well.

Here in this article the investigation of [1] has been continued by the general forms of the partial results found in [1]. The general forms of quantum-mechanical and covariant equations for arbitrary spin together with the general description of the arbitrary spin field formalism are presented. The corresponding relativistic quantum mechanics of arbitrary spin is given as the system of axioms. The ignored in [1] partial example of the spin s=(0,0) particle-antiparticle doublet is considered. Moreover, the study of the place of the results given in [1] among the results of other authors is started. The review of the different investigations in the area of RCQM is given and the brief analysis of the existing approaches to the field theory of arbitrary spin is initiated. 

Note that in the Dirac model [2, 3] the quantum-mechanical interpretation is not evident. It has been demonstrated in [1] that the quantum-mechanical interpretation is much more clear in the Foldy--Wouthuysen (FW) model [4, 5]. Nevertheless, the complete quantum-mechanical picture is possible only in the framework of RCQM. This assertion is one of the main conclusions proved in [1].  

The relativistic quantum mechanics under consideration is called \textit{canonical} due to three main reasons. (i) The model under consideration has direct link with nonrelativistic quantum mechanics based on nonrelativistic Schr$\mathrm{\ddot{o}}$dinger equation. The principles of heredity and correspondence with other models of physical reality leads directly to nonrelativistic Schr$\mathrm{\ddot{o}}$dinger quantum mechanics. (ii) The FW model is already called by many authors as the canonical representation of the Dirac equation or a canonical field model, see, e. g., the paper [5]. And the difference between the field model given by FW and the RCQM is minimal -- in corresponding equations it is only the presence and absence of beta matrix. (iii) The list of relativistic quantum-mechanical models is long. The Dirac model and the FW model are called by the "`old"' physicists as the relativistic quantum mechanics as well (one of my tasks in this paper is to show in visual and demonstrative way that these models have only weak quantum-mechanical interpretation). Further, the fractional relativistic quantum mechanics and the proper-time relativistic quantum mechanics can be listed (recall matrix formulation by W. Heisenberg, Feynman's sum over path's quantum theory, many-worlds interpretation by H. Everett), etc. Therefore, in order to avoid a confusion the model under consideration must have its proper name. Due to the reasons (i)--(iii) the best name for it is RCQM.

The general and fundamental goals in [1] and here are as follows: (i) visual and demonstrative generalization of existing RCQM for the case of arbitrary spin, (ii) more complete formulation of this model on axiomatic level (on the test example of spin s=(1/2,1/2) particle-antiparticle doublet), (iii) vertical and horizontal links between the three different models of physical reality: relativistic quantum mechanics of arbitrary spin in canonical form, canonical (FW type) field theory of any spin, locally covariant (Dirac and Maxwell type) field theory of arbitrary spin.

All results of [1] are confirmed here and consideration of some important details is improved.

The concepts, definitions and notations here are the same as in [1].
For example, in the Minkowski space-time

\begin{equation}
\label{eq1}
\mathrm{M}(1,3)=\{x\equiv(x^{\mu})=(x^{0}=t, \,
\overrightarrow{x}\equiv(x^{j}))\}; \quad \mu=\overline{0,3}, \, j=1,2,3,
\end{equation}
\vskip3mm

\noindent $x^{\mu}$ denotes the Cartesian (covariant)
coordinates of the points of the physical space-time in the
arbitrary-fixed inertial reference frame (IRF). We use the system of units with $\hbar=c=1$. The metric tensor is given by

\begin{equation}
\label{eq2}
g^{\mu\nu}=g_{\mu\nu}=g^{\mu}_{\nu}, \, \left(g^{\mu}_{\nu}\right)=\mathrm{diag}\left(1,-1,-1,-1\right); \quad x_{\mu}=g_{\mu\nu}x^{\mu},
\end{equation}
\vskip3mm

\noindent and summation over the twice repeated indices  is implied.

Thus, in section 2 the Dirac's comment of the square-root operator equation is considered.

In section 3 the additional to [1] comments of the Foldy's contribution are given.

Section 4 contains the RCQM status quo review.

Section 5 contains the equations for arbitrary spin status quo review.

In section 6 the axioms of the RCQM of arbitrary spin are presented. The axiom on the Clifford--Dirac algebra is given in details.

In section 7 the general description of the arbitrary spin field theory is given. The partial examples of the spin s=(0,0) and spin s=(3/2,3/2) particle-antiparticle doublets are visualized.

In section 8 different ways of the Dirac equation derivation are reviewed. The place of our derivations is determined. 

Section 9 contains the brief discussion of interaction in the models under consideration.

Section 10 contains application to the discussion around the antiparticle negative mass.

Section 11 contains discussions and conclusions.

\vskip 0.3cm

\begin{center}
\textbf{Section 2. Dirac's comment}
\end{center}

\vskip 0.3cm

Note that the square-root operator equation, which is the main equation of RCQM, has been rejected by Dirac. In his consideration in [3] (chapter 11, section 67) of the main steps of [2] Dirac discussed this equation. His comment was as follows.

<<\textit{Let us consider first the case of the motion of an electron in the absence of an electromagnetic field, so that the problem is simply that of the free particle, as dealt with in \S 30, with the possible addition of internal degrees of freedom. The relativistic Hamiltonian provided by classical mechanics for this system is given by equation 23 of \S 30, and leads to the wave equation}

$$\left\{p_{0}-\left(m^{2}c^{2}+p^{2}_{1}+p^{2}_{2}+p^{2}_{3}\right)^{\frac{1}{2}}\right\}\psi=0, \quad \quad (5)$$

\noindent \textit{where the $p$'s are interpreted as operators in accordance with (4). Equation (5), although it takes into account the relation between energy and momentum required by relativity, is yet unsatisfactory from the point of view of relativistic theory, because it is very unsymmetrical between $p_{0}$ and the other $p$'s, so much so that one cannot generalize it in a relativistic way to the case when there is a field present. We must therefore look for a new wave equation.}>>

Today another reason of Dirac's rejection of the square-root operator equation is evident as well. The operations with the Dirac's Hamiltonian are too much easier than the operations with the pseudo-differential Hamiltonian of the equation

\begin{equation}
\label{eq3}
i\partial_{t}f(x)=\sqrt{m^{2}-\Delta}f(x)
\end{equation}

\noindent for the N-component wave function
\begin{equation}
\label{eq4}
f\equiv \mathrm{column}(f^{1},f^{2},...,f^{\mathrm {N}}), \quad \mathrm {N}=2s+1,
\end{equation}

\noindent in the case of particle singlet and for the M-component wave function ($\mathrm {M}=2\mathrm {N}=2(2s+1)$) in the case of particle-antiparticle doublet. Note that namely (3) is the equation of motion in the RCQM of arbitrary spin, see [1] for the details.

Nevertheless, today, contrary to the year 1928, the definition of the pseudo-differential (non-local) operator

\begin{equation}
\label{eq5}
\widehat{\omega} \equiv \sqrt{\widehat{\overrightarrow{p}}^{2} + m^2} =\sqrt { - \Delta + m^2}\geq m>0, \quad \widehat{\overrightarrow{p}}\equiv(\widehat{p}^{j})=-i\nabla, \quad \nabla\equiv(\partial_{\ell}),
\end{equation}

\noindent is well known. The action of the operator (5) in the coordinate representation (see, e. g. [6]) is given by

\begin{equation}
\label{eq6}
\widehat{\omega}f(t,\overrightarrow{x}) =\int d^{3}yK(\overrightarrow{x}-\overrightarrow{y})f(t,\overrightarrow{y}),
\end{equation}

\noindent where the function $K(\overrightarrow{x}-\overrightarrow{y})$ has the form $K(\overrightarrow{x}-\overrightarrow{y})=-\frac{2m^{2}K_{2}(m\left|\overrightarrow{x}-\overrightarrow{y}\right|)}{(2\pi)^{2}\left|\overrightarrow{x}-\overrightarrow{y}\right|^{2}}$ and $K_{\nu}(z)$ is the modified Bessel function (Macdonald function), $\left|\overrightarrow{a}\right|$ designates the norm of the vector $\overrightarrow{a}$.

Further, the following integral form

\begin{equation}
\label{eq7}
(\widehat{\omega}f)(t,\overrightarrow{x})=\frac{1}{(2\pi)^{\frac{3}{2}}}\int d^{3}k e^{i\overrightarrow{k}\overrightarrow{x}} \widetilde{\omega}\widetilde{f}(t,\overrightarrow{k}), \quad \widetilde{\omega}\equiv \sqrt{\overrightarrow{k}^{2}+m^{2}}, \quad \widetilde{f}\in \widetilde{\mathrm{H}}^{3,4},
\end{equation}

\noindent of the operator $\widehat{\omega}$ is used often, see, e. g., [5, 7], where $f$ and $\widetilde{f}$ are linked by the 3-dimensional Fourier transformations

\begin{equation}
\label{eq8}
f(t,\overrightarrow{x})=\frac{1}{(2\pi)^{\frac{3}{2}}}\int d^{3}k e^{i\overrightarrow{k}\overrightarrow{x}}\widetilde{f}(t,\overrightarrow{k})\Leftrightarrow \widetilde{f}(t,\overrightarrow{k})=\frac{1}{(2\pi)^{\frac{3}{2}}}\int d^{3}k e^{-i\overrightarrow{k}\overrightarrow{x}}\widetilde{f}(t,\overrightarrow{x}),
\end{equation}

\noindent (in (8) $\overrightarrow{k}$ belongs to the spectrum $\mathrm{R}^{3}_{\vec{k}}$ of the operator $\widehat{\overrightarrow{p}}$, and the parameter $t\in (-\infty,\infty)\subset\mathrm{M}(1,3)$).

Note that the space of states (55) in [1] is invariant with respect to the Fourier transformation (8). Therefore, both $\overrightarrow{x}$-realization (55) in [1] and $\overrightarrow{k}$-realization $\widetilde{\mathrm{H}}^{3,4}$ for the doublet states space are suitable for the purposes of our consideration. In the $\overrightarrow{k}$-realization the Schr$\mathrm{\ddot{o}}$dinger--Foldy equation has the algebraic-differential form

\begin{equation}
\label{eq9}
i\partial_{t}\widetilde{f}(t,\overrightarrow{k})=\sqrt { \overrightarrow{k}^{2} + m^2}\widetilde{f}(t,\overrightarrow{k}); \quad \overrightarrow{k}\in\mathrm{R}^{3}_{\vec{k}}, \quad \widetilde{f}\in \widetilde{\mathrm{H}}^{3,4}.
\end{equation}

\noindent Below in the places, where misunderstanding is impossible, the symbol "tilde" is omitted.

Thus, today on the basis of above given definitions the difficulties, which stopped Dirac in 1928, can be overcome.   

\vskip 0.3cm

\begin{center}
\textbf{Section 3. Foldy's contribution}
\end{center}

\vskip 0.3cm

The name of the person, whose contribution in the theoretical model based on the equation (3) was decisive, is Leslie Lawrance Foldy (1919--2001). 

His interesting biography is presented in [8]. \textit{<<Les was born in Sabinov, Czechoslovakia, on 26 October 1919, into a family with Hungarian roots. His parents named him Laszlo F$\mathrm{\ddot{o}}$ldi. In the turbulent times following World War I, he immigrated with his parents to the US in 1921. His father changed the family's last name and Les's first name; Les later added his middle name, unaware of its more common spellings.>>} 

The first step of Foldy's contribution is visualization of the quantum mechanical interpretation of the Dirac equation on the basis of transformation to the canonical (quantum-mechanical) representation [4]. This transformation was suggested together with the Netherlander Siegfried (Sieg) Wouthuysen (pronounced Vout'-high-sen). In this FW representation of the Dirac equation the quantum-mechanical interpretation is much more clear. Nevertheless, the direct and evident quantum-mechanical interpretation of the spin s=(1/2,1/2) particle-antiparticle doublet can be fulfilled only within the framework of the RCQM: the start was given in [5], see also the consideration in [1]. 

In our investigations we always marked the role of L. Foldy. Taking into account the L. Foldy's contribution in the construction of RCQM and his proof of the principle of correspondence between RCQM and non-relativistic quantum mechanics, we propose [9, 10] and [1] \textit{to call the $N$-component equation (3) as the Schr$\mathrm{\ddot{o}}$dinger-Foldy equation}. Nevertheless, some results of [5] still were missed in the review [1]. Note here that equation (3), which is a direct sum of one component spinless Salpeter equations [11], has been introduced in the formula (21) of [5]. Furthermore, note here that the Poincar$\mathrm{\acute{e}}$ group representation generators (12), (13) in [1] are known from the formulae (B-25)--(B-28) of the L. Foldy's paper [5].

Other results of L. Foldy are already considered in [1].

\vskip 0.3cm
\vskip 0.3cm

\begin{center}
\textbf{Section 4. Brief review of the relativistic canonical quantum mechanics status quo}
\end{center}

\vskip 0.3cm

Contrary to the times of papers [2, 4, 5, 11], the RCQM today is enough approbated and generally accepted theory. The spinless Salpeter equation has been introduced in [11]. The allusion on the RCQM and the first steps are given in [5], where the Salpeter equation for the 2s+1-component wave function was considered and the cases of s=1/2, s=1 were presented as an examples. In [12] Laslo Foldy continued his investigations [5] by the consideration of the relativistic particle systems with interaction. The interaction was introduced by the specific group-theoretical method.

After that in the RCQM were developed both the construction of mathematical foundations and the solution of concrete quantum-mechanical problems for different potentials. Some mathematical foundations and spectral theory of pseudo-differential operator $\sqrt{\overrightarrow{p}^{2}+m^{2}}-Ze^{2}/r$ were given in [13--16]. The application of the RCQM to the quark-antiquark bound state problem can be found in [17, 18]. The numerical solutions of the RCQM equation for arbitrary confining potentials were presented in [18]. In [19] the spinless Salpeter equation for the N particle system of spinless bosons in  gravitational interaction was applied. In [20] a lower bound on the maximum mass of a boson star on the basis of the Hamiltonian  $\sqrt{\overrightarrow{p}^{2}+m^{2}}-\alpha/r$ has been calculated. In [21] results calculated by the author with the spinless Salpeter equation are compared with those obtained from Schrodinger's equation for heavy-quark systems, heavy-light systems, and light-quark systems. In each case the Salpeter energies agree with experiment substantially better than the Schrodinger energies. The paper [22] deal with an investigation of the exact numerical solutions. The spinless Salpeter equation with the Coulomb potential is solved exactly in momentum space and is shown to agree very well with a coordinate-space calculation. In [23, 24] the problem of spectrum of energy eigenvalues calculations on the basis of the spinless Salpeter equation is considered. The spinless relativistic Coulomb problem is studied. It was shown how to calculate, by some special choices of basis vectors in the Hilbert space of solutions, for the rather large class of power-law potentials, at least upper bounds on these energy eigenvalues. The authors of [23, 24] proved that for the lowest-lying levels, this may be done even analytically. In the paper [25] the spinless Salpeter equation was rewritten into integral and integro-differential equations. Some analytical results concerning the spinless Salpeter equation and the action of the square-root operator have been presented. Further, in [26] F. Brau constructed an analytical solution of the one-dimensional spinless Salpeter equation with a Coulomb potential supplemented by a hard core interaction, which keeps the particle in the $x$ positive region. In the context of RCQM based on the spinless Salpeter equation it was shown [27] how to construct a large class of upper limits on the critical value, $g^{(\ell)}_{c}$, of the coupling constant, $g$, of the central potential, $V(r)=-gv(r)$. In [28] a lower bounds on the ground state energy, in one and three dimensions, for the spinless Salpeter equation applicable to potentials, for which the attractive parts are in $\mathrm{L}^{\mathrm {p}}(\mathrm{R}^{\mathrm {n}})$ for some p > n (n = 1 or 3), are found. An extension to confining potentials, which are not in $\mathrm{L}^{\mathrm {p}}(\mathrm{R}^{\mathrm {n}})$, is also presented. In the paper [29], the authors used the theory of fractional powers of linear operators to construct a general (analytic) representation theory for the square-root energy operator $\gamma^{0}\sqrt{\overrightarrow{p}^{2}+m^{2}}+V$ of FW canonical field theory, which is valid for all values of the spin. The example of the spin 1/2 case, considering a few simple yet solvable and physically interesting cases, is presented in details in order to understand how to interpret the operator. Note that corresponding results for the RCQM can be found from the FW canonical field theory results [29] with the help of our transformation (see the section 9 in [1]). Using the momentum space representation, the authors of [30] presented an analytical treatment of the one-dimensional spinless Salpeter equation with a Coulomb interaction. The exact bound-state energy equation was determined. The results obtained were shown to agree very well with exact numerical calculations existing in the literature. In [31] an exact analytical treatment of the spinless Salpeter equation with a one dimensional Coulomb interaction in the context of quantum mechanics with modified Heisenberg algebra implying the existence of a minimal length was presented. The problem was tackled in the momentum space representation. The bound-state energy equation and the corresponding wave functions were exactly obtained. The probability current for a quantum spinless relativistic particle was introduced [6] based on the Hamiltonian dynamics approach using the spinless Salpeter equation. The correctness of the presented formalism was illustrated by examples of exact solutions to the spinless Salpeter equation including the new ones. Thus, in [6] the partial wave packet solutions of this equation have been presented: the solutions for free massless and massive particle on a line, for massless particle in a linear potential, plane wave solution for a free particle (these solution is given here in formula (8) for N-component case), the solution for free massless particle in three dimensions. Further, in the paper [32] other time dependent wave packet solutions of the free spinless Salpeter equation are given. Taking into account the relation of such wave packets to the L$\mathrm{\acute{e}}$vy process the spinless Salpeter equation (in one dimensional space-time) is called in [32] as the L$\mathrm{\acute{e}}$vy-Schr$\mathrm{\ddot{o}}$dinger equation. The several examples of the characteristic behavior of such wave packets have been shown, in particular of the multimodality arising in their evolutions: a feature at variance with the typical diffusive unimodality of both the corresponding L$\mathrm{\acute{e}}$vy process densities and usual Schr$\mathrm{\ddot{o}}$dinger wave functions. A generic upper bound is obtained [33] for the spinless Salpeter equation with two different masses. Analytical results are presented for systems relevant for hadronic physics: Coulomb and linear potentials when a mass is vanishing. A detailed study for the classical and the quantum motion of a relativistic massless particle in an inverse square potential has been presented recently in [34]. The quantum approach to the problem was based on the exact solution of the corresponding spinless Salpeter equation for bound states. Finally, in [38] the connection between the classical and the quantum descriptions via the comparison of the associated probability densities for momentum has been made. The goal of the recent paper [35] is a comprehensive analysis of the intimate relationship between jump-type stochastic processes (e. g. L$\mathrm{\acute{e}}$vy flights) and nonlocal (due to integro-differential operators involved) quantum dynamics. in [35] a special attention is paid to the spinless Salpeter (here, $m\geq 0$) equation and the evolution of various wave packets, in particular to their radial expansion in 3D. Foldy's synthesis of <<covariant particle equations>> is extended to encompass free Maxwell theory, which however is devoid of any <<particle>> content. Links with the photon wave mechanics are explored. The authors of [35] takes into account our results [9] presented also in more earlier preprint, see the last reference in [35].

In the papers [9, 10], where we started our investigations in RCQM, this relativistic model for the test case of the spin s=(1/2,1/2) particle-antiparticle doublet is formulated. In [9], this model is considered as the system of the axioms on the level of the von Neumann monograph [36], where the mathematically well-defined consideration of the nonrelativistic quantum mechanics was given. Furthermore, in [9, 10] the operator link between the spin s=(1/2,1/2) particle-antiparticle doublet RCQM and the Dirac theory is given and Foldy's synthesis of <<covariant particle equations>> is extended to the start from the RCQM of the spin s=(1/2,1/2) particle-antiparticle doublet. In [1] the same procedure is fulfilled for the spin s=(1,1), s=(1,0,1,0), s=(3/2,3/2), s=(2,2), s=(2,0,2,0) and spin s=(2,1,2,1) RCQM. The corresponding equations, which follow from the RCQM for the covariant local field theory, are introduced.

Therefore, here and in [1] I am not going to formulate a new relativistic quantum mechanics! The foundations of RCQM based on the spinless Salpeter equation are already formulated in [5, 6, 9--35]. 

\vskip 0.3cm

\begin{center}
\textbf{Section 5. Brief analysis of the covariant equations for an arbitrary spin}
\end{center}

\vskip 0.3cm

One of the goals of [1] is the link between the RCQM of an arbitrary spin and the different approaches to the covariant local field theory of an arbitrary spin. Surely, at least the brief analysis of the existing covariant equations for an arbitrary spin should be presented.  

Note that in [1] and here only the first-order particle and the field equations (together with their canonical nonlocal pseudo-differential representations) are considered. The second order equations (like the Klein--Gordon--Fock equation) are not the subject of this investigation.

Different approaches to the description of the field theory of an arbitrary spin can be found in [5, 37--46]. Here and in [1] only the approach started in [5] is the basis for further application. Other results given in [5, 37--46] are not used here.

Note only some general deficiencies of the known equations for arbitrary spin. The consideration of the partial cases, when the substitution of the fixed value of spin is fulfilled, is not successful in all cases. For example, for the spin s>1 existing equations have the redundant components and should be complemented by some additional conditions. Indeed, the known equations [47, 48] for the spin s=3/2 (and their confirmation in [49]) should be essentially complemented by the additional conditions. The main difficulty in the models of an arbitrary spin is the interaction between the fields of higher-spin. Even the quantization of higher-spin fields generated the questions. These and other deficiencies of the known equations for higher-spin are considered in [50--61] (a brief review of deficiencies see in [59]).

Equations suggested in [1] and here are free of these deficiencies. The start of such consideration is taken from [5], where the main foundations of the RCQM are formulated. In the text of [1] and here the results of [5] are generalized and extended. The operator link between the results of [4] and [5] (between the canonical FW type field theory and the RCQM) is suggested. Note that the cases s=3/2 and s=2 are not presented in [5], especially in explicit demonstrative forms. The results of [1] are closest to the given in [62, 63]. The difference is explained in the section ?? below. 

Even this brief analysis makes us sure in the prospects of the investigations started in [1]. The successful description of the arbitrary spin field models is not the solved problem today.

\vskip 0.3cm

\begin{center}
\textbf{Section 6. Axioms of the relativistic canonical quantum mechanics of an arbitrary spin}
\end{center}

\vskip 0.3cm

The RCQM of the arbitrary spin given in sections 2 and 18 of [1] can be formulated at the level of von Neumann's consideration [36]. The difference is only in relativistic invariance and in the consideration of multicomponent and multidimensional objects.

The partial case of axiomatic formulation is already given in section 7 of [1] at the example of spin s=1/2 particle-antiparticle doublet. The RCQM of the arbitrary spin particle-antiparticle doublet (or particle singlet) can be formulated similarly as the corresponding generalization of this partial case.

Below the brief presentation of the \textit{list of the axioms} is given. Note that some particular content of these axioms is already given in section 2 of [1], where the RCQM of the arbitrary spin particle singlet has been formulated.

\textbf{On the space of states}. The space of states of isolated  arbitrary spin particle singlet in an arbitrarily-fixed inertial frame of reference (IFR) in its  $\overrightarrow{x}$-realization is the Hilbert space

\begin{equation}
\label{eq10}
\mathrm{H}^{3,\mathrm {N}}=\mathrm{L}_{2}(\mathrm{R}^3)\otimes\mathrm{C}^{\otimes \mathrm{N}}=\{f=(f^{\mathrm {N}}):\mathrm{R}^{3}\rightarrow\mathrm{C}^{\otimes \mathrm {N}}; \quad \int d^{3}x|f(t,\overrightarrow{x})|^{2} <\infty\}, \quad \mathrm {N}=2s+1,
\end{equation}

\noindent of complex-valued N-component square-integrable functions of $x\in\mathrm{R}^{3}\subset \mathrm{M}(1,3)$ (similarly, in momentum,  $\overrightarrow{p}$-realization). In (10) $d^{3}x$ is the Lebesgue measure in the space $\mathrm{R}^{3}\subset \mathrm{M}(1,3)$ of the eigenvalues of the position operator $\overrightarrow{x}$ of the Cartesian coordinate of the particle in an arbitrary-fixed IFR. Further, $\overrightarrow{x}$ and $\overrightarrow{p}$ are the operators of canonically conjugated dynamical variables of the spin s=(1/2,1/2) particle-antiparticle doublet, and the vectors $f$, $\tilde{f}$  in $\overrightarrow{x}$- and $\overrightarrow{p}$-realizations are linked by the 3-dimensional Fourier transformation (the variable $t$ is the parameter of time-evolution).

\textbf{The mathematical correctness of the consideration} demands the application of the rigged Hilbert space 

\begin{equation}
\label{eq11}
\mathrm{S}^{3,\mathrm {N}}\equiv \mathrm{S}(\mathrm{R}^{3})\times\mathrm{C}^{\mathrm {N}}\subset\mathrm{H}^{3,\mathrm {N}}\subset\mathrm{S}^{3,\mathrm {N}*}.
\end{equation}

\noindent where the Schwartz test function space $\mathrm{S}^{3,\mathrm {N}}$ is the core (i. e., it is dense both in $\mathrm{H}^{3,\mathrm {N}}$ and in the space $\mathrm{S}^{3,\mathrm {N}*}$ of the N-component Schwartz generalized functions). The space $\mathrm{S}^{3,\mathrm {N}*}$ is conjugated to that of the
Schwartz test functions $\mathrm{S}^{3,\mathrm {N}}$ by the
corresponding topology (see, e. g. [64]).

Such consideration allows us to perform, without any loss of generality, all necessary calculations in the space $\mathrm{S}^{3,\mathrm {N}}$ at the level of correct differential and integral calculus. The more detailed consideration (at the level of [65]) is given in section 2 of [1]. In the case of arbitrary spin particle-antiparticle doublet the dimension of spaces (10), (11) is M=2N=2(2s+1).   

\textbf{On the time evolution of the state vectors}. The time dependence of the state vectors $f\in \mathrm{H}^{3,\mathrm {N}}$  (time $t$ is the parameter of evolution) is given either in the integral form by the unitary operator

\begin{equation}
\label{eq12}
u\left(t_{0},t\right)=\exp \left[-i \widehat{\omega}(t-t_{0})\right]; \quad \widehat{\omega}\equiv\sqrt{-\Delta +m^{2}},
\end{equation}

\noindent (below $t_{0}=t$ is put), or in the differential form by the Schr$\mathrm{\ddot{o}}$dinger--Foldy equation of motion (3) with the wave function (4). In terms of operator (5)--(7) this equation is given by
\begin{equation}
\label{eq13}
(i\partial_{0}- \widehat{\omega})f(x)=0.
\end{equation}

\noindent Note that here the operator $\widehat{\omega}\equiv\sqrt{-\Delta +m^{2}}$  is the relativistic analog of the energy operator (Hamiltonian) of nonrelativistic quantum mechanics. The Minkowski space-time M(1,3) is pseudo Euclidean with metric $g =\mathrm{diag}(+1,-1,-1,-1)$. The step from the particle singlet of arbitrary spin to the corresponding particle-antiparticle doublet is evident.

Thus, for the arbitrary spin particle-antiparticle doublet the system of two N-component equations $(i\partial_{0}- \widehat{\omega})f(x)=0$ and $(i\partial_{0}- \widehat{\omega})f(x)=0$ is used. Therefore, the corresponding Schr$\mathrm{\ddot{o}}$dinger--Foldy equation is given by (13), where the 2N-component wave function is the direct sum of the particle and antiparticle wave functions, respectively. Due to the historical tradition of the physicists the antiparticle wave function is put in the down part of the 2N-column.

The general solution of the Schr$\mathrm{\ddot{o}}$dinger--Foldy equation of motion (13) (in the case of particle-antiparticle arbitrary spin doublet) has the form

\begin{equation}
\label{eq14}
f(x)= \frac{1}{\left(2\pi\right)^{\frac{3}{2}}}\int d^{3}k e^{-ikx} a^{2\mathrm {N}}\left(\overrightarrow{k}\right)\mathrm{d}_{2\mathrm {N}}, \quad  kx\equiv \omega t -\overrightarrow{k}\overrightarrow{x}, \quad \omega \equiv \sqrt{\overrightarrow{k}^{2}+m^{2}},
\end{equation}

\noindent where the orts of the N-dimensional Cartesian basis are given in [1] by the formulae (10).

The action of the pseudo-differential (non-local) operator $\widehat{\omega}\equiv\sqrt{-\Delta +m^{2}}$ is explained in (6), (7).

\textbf{On the fundamental dynamical variables}. The dynamical variable $\overrightarrow{x}\in \mathrm{R}^{3}\subset$M(1,3) (as well as the variable $\overrightarrow{k}\in \mathrm{R}_{\vec{k}}^{3}$) represents the external degrees of freedom of the arbitrary spin particle-antiparticle doublet. The spin $\overrightarrow{s}$ of the particle-antiparticle doublet is the first in the list of the carriers of the internal degrees of freedom. Taking into account the Pauli principle and the fact that experimentally an antiparticle is observed as the mirror reflection of a particle, the operators of the charge sign and the spin of the arbitrary particle-antiparticle doublet are taken in the form

\begin{equation}
\label{eq15} g \equiv -\Gamma_{2\mathrm{N}}^{0}\equiv -\sigma^{3}_{2\mathrm{N}}=\left| {{\begin{array}{*{20}c}
 -\mathrm{I}_{\mathrm{N}} \hfill &  0 \hfill\\
 0 \hfill & \mathrm{I}_{\mathrm{N}}  \hfill\\
 \end{array} }} \right|,
 \quad \overrightarrow{s}_{2\mathrm{N}}= \left| {{\begin{array}{*{20}c}
 \overrightarrow{s}_{\mathrm {N}} \hfill  & 0 \hfill \\
 0 \hfill & -\hat{C}\overrightarrow{s}_{\mathrm {N}}\hat{C} \hfill \\
 \end{array} }} \right|, \quad \mathrm {N}=2s+1,
\end{equation}

\noindent where $\Gamma_{2\mathrm{N}}^{0}$ is the $2\mathrm{N}\times 2\mathrm{N}$ Dirac $\Gamma^{0}$ matrix, $\sigma^{3}_{2\mathrm{N}}$ is the $2\mathrm{N}\times 2\mathrm{N}$ Pauli $\sigma^{3}$ matrix, $\hat{C}$ is the operator of complex conjugation in the form of $\mathrm{N}\times \mathrm{N}$ diagonal matrix, the operator of involution in $\mathrm{H}^{3,2\mathrm{N}}$, and $\mathrm{I}_{\mathrm{N}}$ is $\mathrm{N}\times \mathrm{N}$ unit matrix.

Thus, the spin is given by the generators of SU(2) algebra! 

The spin matrices $\overrightarrow{s}_{2\mathrm{N}}$ (15) satisfy the commutation relations 

\begin{equation}
\label{eq16}
\left[s^{j}_{2\mathrm{N}},s^{\ell}_{2\mathrm{N}}\right]=i\varepsilon^{j \ell n}s^{n}_{2\mathrm{N}}, \quad \varepsilon^{123}=+1,
\end{equation}

\noindent of the algebra of SU(2) group, where $\varepsilon^{j \ell n}$ is the Levi-Civita tensor and $s^{j}=\varepsilon^{j \ell n}s_{\ell n}$ are the Hermitian $2\mathrm{N}\times 2\mathrm{N}$ matrices (15) -- the generators of a 2N-dimensional reducible representation of the spin group SU(2) (universal covering of the SO(3)$\subset$SO(1,3) group).

The Casimir operator for the RCQM representation of SU(2) spin given in (15) has the form

\begin{equation}
\label{eq17} \overrightarrow{s}^{2}_{2\mathrm{N}}=\mathrm{s}(\mathrm{s}+1)\mathrm{I}_{2\mathrm{N}},
\end{equation}

\noindent where $\mathrm{I}_{2\mathrm{N}}$ is $2\mathrm{N}\times 2\mathrm{N}$ unit matrix.

Above in the text of this axiom the case of arbitrary spin \textit{particle-antiparticle doublet} has been considered. For the case of arbitrary spin \textit{particle singlet} the operator of the charge sign is absent. The spin operator $\overrightarrow{s}_{\mathrm{N}}$ is given by the Hermitian $\mathrm{N}\times \mathrm{N}$ matrices -- the generators of a N-dimensional irreducible representation of the spin group SU(2). Therefore, the SU(2) commutation relations in such notations have the explicit form

\begin{equation}
\label{eq18}
\left[s^{j}_{\mathrm{N}},s^{\ell}_{\mathrm{N}}\right]=i\varepsilon^{j \ell n}s^{n}_{\mathrm{N}}, \quad \varepsilon^{123}=+1,
\end{equation}

\noindent and the corresponding Casimir operator is given by

\begin{equation}
\label{eq19} \overrightarrow{s}^{2}_{\mathrm{N}}=\mathrm{s}(\mathrm{s}+1)\mathrm{I}_{\mathrm{N}}.
\end{equation}

The validity of these assertions is proved by numerous partial examples presented in [1].

\textbf{On the external and internal degrees of freedom}. The coordinate $\overrightarrow{x}$ (as an operator in $\mathrm{H}^{3,2\mathrm{N}})$ is an analog of the discrete index of generalized coordinates $q\equiv(q_{1}, q_{2}, ...)$ in non-relativistic quantum mechanics of the finite number degrees of freedom. In other words the coordinate $\overrightarrow{x}\in \mathrm{R}^{3}\subset$M(1,3) is the continuous carrier of the external degrees of freedom of a multiplet (the similar consideration was given in [66]). The coordinate operator together with the operator $\widehat{\overrightarrow{p}}$ determines the operator $m_{ln}=x_{l}\widehat{p}_{n}-x_{n}\widehat{p}_{l}$ of an orbital angular momentum, which also is connected with the external degrees of freedom.

However, the RCQM doublet has the additional characteristics such as the spin operator $\overrightarrow{s}$ (15), which is the carrier of the internal degrees of freedom of this multiplet. The set of generators $(\widehat{p}_{\mu}, \widehat{j}_{\mu\nu})$ (formulae (12), (13) in [1]) of the main dynamical variables (formulae (73) in [1]) of the doublet are the functions of the following basic set of 9 functionally independent operators

\begin{equation}
\label{eq20}
\overrightarrow{x}=(x^{j}), \, \widehat{\overrightarrow{p}}=(\widehat{p}^{j}), \, \overrightarrow{s}_{2\mathrm{N}} \equiv\left(s^{j}_{2\mathrm{N}}\right)=\left(s_{23},s_{31},s_{12}\right).
\end{equation}

Note that spin $\overrightarrow{s}_{2\mathrm{N}}$ (15) commutes both with $(\overrightarrow{x},\widehat{\overrightarrow{p}})$ and with the operator $i\partial_{t}-\sqrt { - \Delta + m^2}$ of the Schr$\mathrm{\ddot{o}}$dinger--Foldy equation (13). Thus, for the free doublet the external and internal degrees of freedom are independent. Therefore, 9 operators (20) in $\mathrm{H}^{3,2\mathrm{N}}$, which have the univocal physical sense, are the \textit{generating} operators not only for the 10 $\mathcal{P}$ generators $(\widehat{p}_{\mu}, \widehat{j}_{\mu\nu})$ (12), (13) of [1], but also for other operators of any experimentally observable quantities of the doublet.

\textbf{On the algebra of observables}. Using the operators of canonically conjugated coordinate $\overrightarrow{x}$ and momentum $\overrightarrow{p}$ (where $\left[x^{j},\widehat{p}^{\ell}\right]=i\delta^{j \ell}, \quad \left[x^{j},x^{\ell}\right]=\left[\widehat{p}^{j},\widehat{p}^{\ell}\right]=0,$) in $\mathrm{H}^{3,2\mathrm{N}}$, being completed by the operators $\overrightarrow{s}_{2\mathrm{N}}$ and $g$ (15), we construct the algebra of observables as the Hermitian functions of 10 ($\overrightarrow{x}, \, \overrightarrow{p}, \, \overrightarrow{s}_{2\mathrm{N}}, \, -\Gamma_{2\mathrm{N}}^{0}$) generating elements of the algebra.

\textbf{On the relativistic invariance of the theory}. The main assertions of this axiom are considered already in section 2 of [1]. Recall briefly that the relativistic invariance of the RCQM (implementation of the special relativity) is ensured by the proof of the invariance of the Schr$\mathrm{\ddot{o}}$dinger--Foldy equation (13) with respect to the unitary representation of the \textit{universal covering} $\mathcal{P}\supset\mathcal{L}$=SL(2,C) of the proper ortochronous Poincar$\mathrm{\acute{e}}$ group $\mbox{P}_ + ^
\uparrow = \mbox{T(4)}\times )\mbox{L}_ + ^ \uparrow  \supset \mbox{L}_ + ^ \uparrow$ (here $\mathcal{L}$ = SL(2,C) is the universal covering of proper ortochronous Lorentz group $\mbox{L}_ + ^ \uparrow $). Another important assertion of [1] is the possibility to use the \textit{non-covariant objects and non-covariant Poincar$\mathrm{\acute{e}}$ generators}. Not a matter of fact that non-covariant objects such as the Lebesgue measure $d^{3}x$ and non-covariant generators of algebras are explored, the model of RCQM of arbitrary spin is a relativistic invariant. The three step proof of this assertion is given in section 2 of [1] as well. 

In addition to [1] note that together with the generators (12), (13) of [1] another set of 10 operators commutes with the operator of equation (13), satisfies the commutation relations ((11) of [1]) of the Lie algebra of Poincar$\mathrm{\acute{e}}$ group $\mathcal{P}$, and, therefore, can be chosen as the Poincar$\mathrm{\acute{e}}$ symmetry of the model under consideration. This second set is given by the generators $\widehat{p}^{0}, \, \widehat{p}^{\ell}$ from (12) in [1] together with the orbital parts of the generators $\widehat{j}^{\ell
n}, \, \widehat{j}^{0 \ell}$ from (13) in [1]. Thus, this second set of Poincar$\mathrm{\acute{e}}$ generators is given by

\begin{equation}
\label{eq21}
\widehat{p}_{0}=\widehat{\omega}\equiv \sqrt{-\Delta+m^{2}}, \quad \widehat{p}_{\ell}=i\partial_{\ell}, \quad
\widehat{m}_{\ell n}=x_{\ell}\widehat{p}_{n}-x_{n}\widehat{p}_{\ell}, \quad
\widehat{m}_{0 \ell}=-\widehat{m}_{\ell 0}=t\widehat{p}_{\ell}-\frac{1}{2}\left\{x_{\ell},\widehat{\omega}\right\}.
\end{equation}

\noindent Note that in the case s=0 only generators (21) form the Poincar$\mathrm{\acute{e}}$ symmetry.

Next comment to the consideration of [1] is as follows. The expression

\begin{equation}
\label{eq22} (a,\varpi)\rightarrow U(a,\varpi)=\exp
(-ia^{0}\widehat{p}_{0}-i\overrightarrow{a}\widehat{\overrightarrow{p}}-\frac{i}{2}\varpi^{\mu\nu}\widehat{j}_{\mu\nu})
\end{equation}

\noindent ((14) in [1]) is well known, but rather formal. In fact the transition from a Lie algebra to a finite group transformations in the case of non-Lie operators is a rather non-trivial action. The mathematical justification of (22) can be fulfilled in the framework of Schwartz test function space and will be given in next special publication.

Note that the modern definition of $\mathcal{P}$ invariance (or $\mathcal{P}$ symmetry) of the equation of motion (13) in $\mathrm{H}^{3,\mathrm{N}}$ is given by the following assertion, see, e. g. [67]. \textit{The set} $\mathrm{F}\equiv\left\{f\right\}$ \textit{of all possible solutions of the equation (13) is invariant with respect to the} $\mathcal{P}^{\mathrm{f}}$-\textit{representation of the group} $\mathcal{P}$, \textit{if for arbitrary solution} $f$ \textit{and arbitrarily-fixed parameters} $(a,\varpi)$ \textit{the assertion}

\begin{equation}
\label{eq23}
(a,\varpi)\rightarrow U(a,\varpi)\left\{f\right\}=\left\{f\right\}\equiv\mathrm{F}
\end{equation}

\noindent \textit{is valid}.

\textbf{On the dynamic and kinematic aspects of the relativistic invariance}. Consider briefly some detalizations of the relativistic invariance of the Schr$\mathrm{\ddot{o}}$dinger--Foldy equation (13). Note that for the free particle-antiparticle doublet of arbitrary spin the equation (13) has one and the same explicit form in arbitrary-fixed IFR (its set of solutions is one and the same in every IFR). Therefore, the algebra of observables and the conservation laws (as the functionals of the free particle-antiparticle doublet states) have one and the same form too. This assertion explains the dynamical sense of the $\mathcal{P}$ invariance (the invariance with respect to the dynamical symmetry group $\mathcal{P}$).

Another, kinematical, aspect of the $\mathcal{P}$ invariance of the RQCM model has the following physical sense. Note at first that any solution of the Schr$\mathrm{\ddot{o}}$dinger--Foldy equation (13) is determined by the concrete given set of the amplitudes $\left\{A\right\}$. It means that if $f$ with the fixed set of amplitudes $\left\{A\right\}$ is the state of the doublet in some arbitrary IFR, then for the observer in the $(a, \, \varpi)$-transformed $\mathrm{IFR}^{\prime}$ this state $f^{\prime}$ is determined by the amplitudes $\left\{A^{\prime}\right\}$. The last ones are received from the given $\left\{A\right\}$ by the unitary  $\mathcal{P}^{\mathrm{A}}$ -transformation (22).

\textbf{On the Clifford--Dirac algebra}. This axiom is additional and is not necessary. Nevertheless, such axiom is very useful for the dimensions, where the $\Gamma$ matrices exist.

Application of the Clifford--Dirac algebra is the useful method of calculations in RCQM. Three different definitions of the Clifford algebra and their equivalence are considered in [68]. In different approaches to the relativistic quantum mechanics the matrix representation of the Clifford algebra in terms of the Dirac gamma matrices is used. This representation is called the Clifford--Dirac algebra.

For our purposes the anticommutation relations of the Clifford--Dirac algebra generators are taken in the general form 

\begin{equation}
\label{eq24}
\Gamma_{2\mathrm{N}}^{\bar{\mu}}\Gamma_{2\mathrm{N}}^{\bar{\nu}}+\Gamma_{2\mathrm{N}}^{\bar{\nu}}\Gamma_{2\mathrm{N}}^{\bar{\mu}}=2g^{\bar{\mu}\bar{\nu}}; \quad \bar{\mu},\bar{\nu} = \overline{0,4}, \quad (g^{\bar{\mu}\bar{\nu}})=(+----),
\end{equation}

\noindent where $\Gamma_{2\mathrm{N}}^{\bar{\mu}}$ are the $2\mathrm{N}\times 2\mathrm{N}$ Dirac $\Gamma^{\bar{\mu}}$ matrices ($2\mathrm{N}\times 2\mathrm{N}$ generalization of the Dirac $4 \times 4$ $\gamma$ matrices), $\Gamma_{2\mathrm{N}}^{4}\equiv \Gamma_{2\mathrm{N}}^{0}\Gamma_{2\mathrm{N}}^{1}\Gamma_{2\mathrm{N}}^{2}\Gamma_{2\mathrm{N}}^{3}$. Here and in our publications (see, e. g. the last years articles [69--73]) we use the $\gamma^{4}\equiv \gamma^{0}\gamma^{1}\gamma^{2}\gamma^{3}$ matrix instead of the $\gamma^{5}$ matrix of other authors. Our $\gamma^{4}$ is equal to $i\gamma^{5}_{\mathrm{standard}}$. Notation $\gamma^{5}$ is used in [69--73] for a completely different matrix $\gamma^{5}\equiv \gamma^{1}\gamma^{3}\hat{C}$. As well as the element of the algebra $\gamma^{4}\equiv \gamma^{0}\gamma^{1}\gamma^{2}\gamma^{3}$ is dependent the algebra basis is formed by 4=1+3 independent elements. Therefore, such Clifford algebra over the field of complex numbers is denoted Cl$^{\texttt{C}}$(1,3) and the dimension of the algebra is $2^{4}=16$.

Note that relations (24) are valid only for the dimensions, where the $\Gamma_{2\mathrm{N}}$ are defined (where the matrix representation of the Clifford algebra exists). Therefore, the relations (24) can be useful not for all multiplets of RCQM. Nevertheless, existing relations (24) are very useful in order to operate with spins, with standard FW transformation and in order to formulate the FW transformation generalizations for the dimensions 2(2s+1)>4.      

It has been explained in [1] (and in [69--73] in details) that the Clifford--Dirac algebra should be introduced into consideration in the FW representation [4] of the spinor field. The reasons are as follows. Part of the Clifford--Dirac algebra operators are directly related to the spin 1/2 doublet operators $(s^{1}\equiv \frac{1}{2}\gamma^{2}\gamma^{3}, \, s^{2}\equiv \frac{1}{2}\gamma^{3}\gamma^{1}, \, s^{3}\equiv \frac{1}{2}\gamma^{1}\gamma^{2})$ (in the anti-Hermitian form). In the FW representation for the spinor field [4] these spin operators commute with the Hamiltonian and with the operator of the FW equation of motion $i\partial _{0}-\gamma^{0}\widehat{\omega}$. In the Pauli--Dirac representation these operators do not commute with the Dirac equation operator. Only the sums of the orbital operators and such spin operators commute with the Diracian. So \textit{if we want to relate the orts $\gamma^{\mu}$ of the Clifford--Dirac algebra with the actual spin we must introduce this algebra into the FW representation}.

Here in general N dimensional formalism of RCQM the situation is similar. The anticommutation relations of the Clifford--Dirac algebra generators (24) and corresponding $\Gamma_{2\mathrm{N}}$ matrices must be introduced in the FW representation. Therefore, in order to apply (24) in RCQM one must transform the $\Gamma_{2\mathrm{N}}$ matrices from the FW representation into the RCQM representation. Corresponding operator transformation is given by

\begin{equation}
\label{eq25}
v_{2\mathrm{N}} = \left| {{\begin{array}{*{20}c}
 \mathrm{I}_{\mathrm{N}} \hfill &  0 \hfill\\
 0 \hfill & \hat{C}\mathrm{I}_{\mathrm{N}}  \hfill\\
 \end{array} }} \right|, \quad v_{2\mathrm{N}}^{-1}= v_{2\mathrm{N}}^{\dag}= v_{2\mathrm{N}}, \quad v_{2\mathrm{N}}v_{2\mathrm{N}}=\mathrm{I}_{2\mathrm{N}}, \quad \mathrm{N}=2s+1,
\end{equation}
\vskip3mm 

\noindent where $\hat{C}\mathrm{I}_{\mathrm{N}}$ is the $\mathrm{N} \times \mathrm{N}$ operator of complex conjugation. Indeed, the operator (25) translates any operator from canonical field FW representation into the RCQM representation and vice versa:

\begin{equation}
\label{eq26} v_{2\mathrm{N}}\hat{q}_{\mathrm{cf}}^{\mathrm{anti-Herm}}v_{2\mathrm{N}}
= \hat{q}_{\mathrm{qm}}^{\mathrm{anti-Herm}}, \quad v_{2\mathrm{N}}\hat{q}_{\mathrm{qm}}^{\mathrm{anti-Herm}}v_{2\mathrm{N}}
= \hat{q}_{\mathrm{cf}}^{\mathrm{anti-Herm}}.
\end{equation}

\noindent Here $\hat{q}_{\mathrm{qm}}^{\mathrm{anti-Herm}}$ is an arbitrary operator from the RCQM of the 2N-component particle-antiparticle doublet in the anti-Hermitian form, e. g., the operator $(\partial_{0}+i\widehat{\omega})$ of equation of motion (13), the operator of spin $\overrightarrow{s}_{2\mathrm{N}}$ (15) taken in anti-Hermitian form, etc.,  $\hat{q}_{\mathrm{cf}}^{\mathrm{anti-Herm}}$ is an arbitrary operator from the canonical field theory of the 2N-component particle-antiparticle doublet in the anti-Hermitian form. Thus, the only warning is that operators here must be taken in anti-Hermitian form, see section 9 in [1] for the details and see [74, 75] for the mathematical correctness of  anti-Hermitian operators application.  

Further, the operator (25) translates 

\begin{equation}
\label{eq27} \phi=v_{2\mathrm{N}} f, \quad f=v_{2\mathrm{N}}\phi,
\end{equation}

\noindent the solution (14) of the Schr$\mathrm{\ddot{o}}$dinger--Foldy equation (13) into the solution 

\begin{equation}
\label{eq28} \phi(x)=
\frac{1}{\left(2\pi\right)^{\frac{3}{2}}}\int
d^{3}k\left[e^{-ikx}a^{\mathrm{N}}(\overrightarrow{k})\mathrm{d}_{\mathrm{N}}+e^{ikx}a^{*\breve{\mathrm{N}}}(\overrightarrow{k})\mathrm{d}_{\breve{\mathrm{N}}}\right],
\end{equation}
$\mathrm{N} =1,2,...,\mathrm{N}, \quad \breve{\mathrm{N}} = \mathrm{N}+1,\mathrm{N}+2,...,2\mathrm{N},$ of the FW equation

\begin{equation}
\label{eq29} (i\partial_{0}-
\Gamma_{2\mathrm{N}}^{0}\widehat{\omega})\phi(x)=0, \quad \Gamma_{2\mathrm{N}}^{0}\equiv \sigma^{3}_{2\mathrm{N}}=\left| {{\begin{array}{*{20}c}
 \mathrm{I}_{\mathrm{N}} \hfill &  0 \hfill\\
 0 \hfill & -\mathrm{I}_{\mathrm{N}}  \hfill\\
 \end{array} }} \right|,
\end{equation}
$\widehat{\omega}\equiv \sqrt{-\Delta+m^{2}},  \quad \mathrm{N}=2s+1,$ and vice versa.

Thus, the transformation (25), (26) translates the matrices $\Gamma_{2\mathrm{N}}^{0}$ and 

\begin{equation}
\label{eq30} \Gamma_{2\mathrm{N}}^{j}=\left| {{\begin{array}{*{20}c}
 0 \hfill & \Sigma_{\mathrm{N}}^{j} \\
 -\Sigma_{\mathrm{N}}^{j} \hfill & 0 \\
\end{array} }} \right|,  \quad j=1,2,3,
\end{equation}
\vskip3mm

\noindent into the RCQM representation

\begin{equation}
\label{eq31}
\bar{\Gamma}_{2\mathrm{N}}^{\bar{\mu}}= v_{2\mathrm{N}}\Gamma_{2\mathrm{N}}^{\bar{\mu}}v_{2\mathrm{N}},
\end{equation}

\noindent where matrices $\bar{\Gamma}_{2\mathrm{N}}^{\bar{\mu}}$ satisfy the anticommutation relations 

\begin{equation}
\label{eq32}
\bar{\Gamma}_{2\mathrm{N}}^{\bar{\mu}}\bar{\Gamma}_{2\mathrm{N}}^{\bar{\nu}}+\bar{\Gamma}_{2\mathrm{N}}^{\bar{\nu}}\bar{\Gamma}_{2\mathrm{N}}^{\bar{\mu}}=2g^{\bar{\mu}\bar{\nu}}; \quad \bar{\mu},\bar{\nu} = \overline{0,4}, \quad (g^{\bar{\mu}\bar{\nu}})=(+----),
\end{equation}

\noindent of the Clifford--Dirac algebra generators as well. In (30) $\Sigma_{\mathrm{N}}^{j}$ are the $\mathrm{N} \times \mathrm{N}$ Pauli matrices. The explicit forms of the RCQM representation of the $\bar{\Gamma}_{2\mathrm{N}}^{\bar{\mu}}$ matrices are given by

\begin{equation}
\label{eq33}
\bar{\Gamma}_{2\mathrm{N}}^{0}= \Gamma_{2\mathrm{N}}^{0}, \quad \bar{\Gamma}_{2\mathrm{N}}^{1}= \Gamma_{2\mathrm{N}}^{1}\widehat{C}, \quad \bar{\Gamma}_{2\mathrm{N}}^{2}= \Gamma_{2\mathrm{N}}^{0}\Gamma_{2\mathrm{N}}^{2}\widehat{C}, \quad \bar{\Gamma}_{2\mathrm{N}}^{3}= \Gamma_{2\mathrm{N}}^{3}\widehat{C}, \quad \bar{\Gamma}_{2\mathrm{N}}^{4}= \Gamma_{2\mathrm{N}}^{0}\Gamma_{2\mathrm{N}}^{4}\widehat{C},
\end{equation}

\noindent where $\widehat{C}$ is the $2\mathrm{N} \times 2\mathrm{N}$ operator of complex conjugation and matrices $\Gamma_{2\mathrm{N}}^{\bar{\mu}}$ are given in (29), (30).

Note that in the terms of $\bar{\Gamma}_{2\mathrm{N}}^{\bar{\mu}}$ matrices (33) the RCQM spin operator (15) has the form

\begin{equation}
\label{eq34}
\overrightarrow{s}=\frac{i}{2}(\bar{\Gamma}_{2\mathrm{N}}^{2}\bar{\Gamma}_{2\mathrm{N}}^{3}, \, \bar{\Gamma}_{2\mathrm{N}}^{3}\bar{\Gamma}_{2\mathrm{N}}^{1}, \, \bar{\Gamma}_{2\mathrm{N}}^{1}\bar{\Gamma}_{2\mathrm{N}}^{2}).
\end{equation}

\noindent Note further that formula (34) is valid for the multiplets of arbitrary dimension but only for the spin s=1/2, whereas the formula (15) is valid for arbitrary spin. Furthermore, the complete analogy between the (34) and the particle-antiparticle spin s=1/2 doublet of arbitrary dimension in the FW representation exists

\begin{equation}
\label{eq35} \overrightarrow{s}_{\mathrm{FW}} =
\frac{i}{2}(\Gamma_{2\mathrm{N}}^{2}\Gamma_{2\mathrm{N}}^{3}, \, \Gamma_{2\mathrm{N}}^{3}\Gamma_{2\mathrm{N}}^{1}, \, \Gamma_{2\mathrm{N}}^{1}\Gamma_{2\mathrm{N}}^{2}).
\end{equation}

It is very useful to consider a wider then Cl$^{\texttt{C}}$(1,3) Clifford--Dirac algebra. In [69--73] such additional algebras have been introduced for the purposes of finding links between the fermionic and bosonic states of the spinor field. New algebra can be formed by the generators of Cl$^{\texttt{C}}$(1,3) together with the generators of the Pauli--Gursey--Ibragimov algebra [76--78].

The main structure elements of such set are given by ($\gamma^{0}, \gamma^{1}, \gamma^{2}, \gamma^{3}, i, \hat{C}\mathrm{I}_{4}$), where $\gamma^{\mu}$ are $4 \times 4$ Dirac matrices in standard representation. It is easy to see that simplest set of the Clifford--Dirac algebra generators can be constructed from these elements in the form ($i\gamma^{0}, i\gamma^{1}, \gamma^{2}, i\gamma^{3}, \hat{C}\mathrm{I}_{4}, i\hat{C}\mathrm{I}_{4}$). Therefore, the $2\mathrm{N} \times 2\mathrm{N}$ matrix generators of the corresponding Clifford--Dirac algebra over the field of real numbers can be found by the simple redefinition

\begin{equation}
\label{eq36} \tilde{\Gamma}_{2\mathrm{N}}^{1}\equiv i\Gamma_{2\mathrm{N}}^{1}, \quad \tilde{\Gamma}_{2\mathrm{N}}^{2}\equiv i\Gamma_{2\mathrm{N}}^{3}, \quad \tilde{\Gamma}_{2\mathrm{N}}^{3}\equiv \widehat{C}\mathrm{I}_{2\mathrm{N}}, \quad \tilde{\Gamma}_{2\mathrm{N}}^{4}\equiv i\widehat{C}\mathrm{I}_{2\mathrm{N}}, \quad \tilde{\Gamma}_{2\mathrm{N}}^{5}\equiv i\Gamma_{2\mathrm{N}}^{0}, \quad \tilde{\Gamma}_{2\mathrm{N}}^{6}\equiv -\Gamma_{2\mathrm{N}}^{2},
\end{equation}
\vskip3mm

\noindent of the matrices ($i\Gamma_{2\mathrm{N}}^{0}, i\Gamma_{2\mathrm{N}}^{1}, \Gamma_{2\mathrm{N}}^{2}, i\Gamma_{2\mathrm{N}}^{3}, \widehat{C}\mathrm{I}_{2\mathrm{N}}, i\widehat{C}\mathrm{I}_{2\mathrm{N}}$), where $\Gamma_{2\mathrm{N}}^{\mu}$ are given in (29), (30).

Matrices (36) together with the matrix $\tilde{\Gamma}_{2\mathrm{N}}^{7}\equiv \tilde{\Gamma}_{2\mathrm{N}}^{1}\tilde{\Gamma}_{2\mathrm{N}}^{2}\tilde{\Gamma}_{2\mathrm{N}}^{3}\tilde{\Gamma}_{2\mathrm{N}}^{4}\tilde{\Gamma}_{2\mathrm{N}}^{5}\tilde{\Gamma}_{2\mathrm{N}}^{6}=\Gamma_{2\mathrm{N}}^{4}$ satisfy the anticommutation relations of the Clifford--Dirac algebra generators in the form 

\begin{equation}
\label{eq37}
\tilde{\Gamma}_{2\mathrm{N}}^{\mathrm{A}}\tilde{\Gamma}_{2\mathrm{N}}^{\mathrm{B}}+\tilde{\Gamma}_{2\mathrm{N}}^{\mathrm{B}}\tilde{\Gamma}_{2\mathrm{N}}^{\mathrm{A}}=2g^{\mathrm{A}\mathrm{B}}; \quad \mathrm{A},\mathrm{B} = \overline{1,7}, \quad (g^{\bar{\mu}\bar{\nu}})=(++++---).
\end{equation}

As well as in (24) among the generators of (37) only the 4+2=6 matrices (36) are independent and form the basis of the algebra. Therefore, the found above algebra over the field of real numbers is defined as Cl$^{\texttt{R}}$(4,2) and the dimension of this algebra is $2^{6}=64$.  

Useful realization of (36), (37) is given in terms of completely anti-Hermitian generators

\begin{equation}
\label{eq38}
\left\{\Gamma_{2\mathrm{N}}^{1},\,\Gamma_{2\mathrm{N}}^{2},\,\Gamma_{2\mathrm{N}}^{3},\,\Gamma_{2\mathrm{N}}^{4}=\Gamma_{2\mathrm{N}}^{0}\Gamma_{2\mathrm{N}}^{1}\Gamma_{2\mathrm{N}}^{2}\Gamma_{2\mathrm{N}}^{3},\,\Gamma_{2\mathrm{N}}^{5}=\Gamma_{2\mathrm{N}}^{1}\Gamma_{2\mathrm{N}}^{3}\widehat{C},\,\Gamma_{2\mathrm{N}}^{6}=i\Gamma_{2\mathrm{N}}^{1}\Gamma_{2\mathrm{N}}^{3}\widehat{C},\,\Gamma_{2\mathrm{N}}^{7}=i\Gamma_{2\mathrm{N}}^{0} \right\},
\end{equation}

\noindent where matrices $\left\{\Gamma_{2\mathrm{N}}^{\mathrm{\mu}}\right\}$ are givem in (29), (30). Matrices (38) obey the anticommutation relations of 64 dimensional Cl$^{\texttt{R}}$(0,6) algebra in the form

\begin{equation}
\label{eq39} \Gamma_{2\mathrm{N}}^{\mathrm{A}} \Gamma_{2\mathrm{N}}^{\mathrm{B}} + \Gamma_{2\mathrm{N}}^{\mathrm{B}}\Gamma_{2\mathrm{N}}^{\mathrm{A}} = -
2\delta^{\mathrm{A}\mathrm{B}},\quad \mathrm{A},\mathrm{B}=\overline{1,7}.
\end{equation}

\noindent Note that here as well only 6 operators are independent generators: $\Gamma_{2\mathrm{N}}^{4}=-i\Gamma_{2\mathrm{N}}^{7}\Gamma_{2\mathrm{N}}^{1}\Gamma_{2\mathrm{N}}^{2}\Gamma_{2\mathrm{N}}^{3}$.

Operators (38) generate also the 28 orts $\alpha^{\bar{\mathrm {A}}\bar{\mathrm {B}}} =
2{s}^{\bar{\mathrm {A}}\bar{\mathrm {B}}}$:

\begin{equation}
\label{eq40}
s^{\bar{\mathrm{A}}\bar{\mathrm{B}}}=\{s^{\mathrm{A}\mathrm{B}}=\frac{1}{4}[\Gamma_{2\mathrm{N}}^{\mathrm{A}},\Gamma_{2\mathrm{N}}^{\mathrm{B}}], \, s^{\mathrm{A}8}=-s^{8\mathrm{A}}=\frac{1}{2}\Gamma_{2\mathrm{N}}^{\mathrm{A}}\},\quad \bar{\mathrm{A}},\bar{\mathrm{B}}=\overline{1,8},
\end{equation}

\noindent where generators $s^{\bar{\mathrm{A}}\bar{\mathrm{B}}}$ satisfy the commutation relations of SO(8) algebra

\begin{equation}
\label{eq41}
[s^{\bar{\mathrm{A}}\bar{\mathrm{B}}},s^{\bar{\mathrm{C}}\bar{\mathrm{D}}}]=
\delta^{\bar{\mathrm{A}}\bar{\mathrm{C}}}s^{\bar{\mathrm{B}}\bar{\mathrm{D}}}
+\delta^{\bar{\mathrm{C}}\bar{\mathrm{B}}}s^{\bar{\mathrm{D}}\bar{\mathrm{A}}}
+\delta^{\bar{\mathrm{B}}\bar{\mathrm{D}}}s^{\bar{\mathrm{A}}\bar{\mathrm{C}}}
+\delta^{\bar{\mathrm{D}}\bar{\mathrm{A}}}s^{\bar{\mathrm{C}}\bar{\mathrm{B}}}.
\end{equation}

\noindent Of course, the algebra SO(8) is considered over the field of real numbers.

As the consequences of the equalities

\begin{equation}
\label{eq42}
\Gamma_{2\mathrm{N}}^{4}\equiv \prod^{3}_{\mu=0}\Gamma_{2\mathrm{N}}^{\mu} \rightarrow \prod^{4}_{\bar{\mu}=0}\Gamma_{2\mathrm{N}}^{\bar{\mu}} = -\mathrm{I},
\end{equation}

\noindent known from the standard Clifford--Dirac algebra Cl$^{\texttt{C}}$(1,3), and the anticommutation relations (39), in Cl$^{\texttt{R}}$(0,6) algebra for the matrices $\Gamma_{2\mathrm{N}}^{\mathrm{A}}$ (38) the following extended equalities are valid:

\begin{equation}
\label{eq43}
\Gamma_{2\mathrm{N}}^{7}\equiv -\prod^{6}_{\underline{\mathrm{A}}=1}\Gamma_{2\mathrm{N}}^{\underline{\mathrm{A}}} \rightarrow \prod^{7}_{\mathrm{A}=1}\Gamma_{2\mathrm{N}}^{\mathrm{A}} = \mathrm{I}, \quad \Gamma_{2\mathrm{N}}^{5}\Gamma_{2\mathrm{N}}^{6}=i.
\end{equation}

The relationship (40) between the Clifford--Dirac algebra Cl$^{\texttt{R}}$(0,6) and algebra SO(8) is similar to the relationship between the standard Clifford--Dirac algebra Cl$^{\texttt{C}}$(1,3) and algebra SO(3,3) found in [79, 80]; for the relationship between Cl$^{\texttt{C}}$(1,3) and SO(1,5) see in [69--73].  

Note that subalgebra SO(6)$\subset$SO(8) is the algebra of invariance of the Dirac equation in the FW representation [4, 5]. The 16 elements of this SO(6) algebra are given by

\begin{equation}
\label{eq44}
\left\{\mathrm{I}, \, \alpha^{\underline{\mathrm{A}} \,\underline{\mathrm{B}}} =
2{s}^{\underline{\mathrm{A}} \, \underline{\mathrm{B}}}\right\},  \quad  \underline{\mathrm{A}},\underline{\mathrm{B}} = \overline{1,6},
\end{equation}

\noindent where

\begin{equation}
\label{eq45}
\left\{s^{\underline{\mathrm{A}} \,\underline{\mathrm{B}}}\right\}=\left\{s^{\underline{\mathrm{A}} \,\underline{\mathrm{B}}}\equiv \frac{1}{4}{\left[\Gamma_{2\mathrm{N}}^{\underline{\mathrm{A}}},\Gamma_{2\mathrm{N}}^{\underline{\mathrm{B}}}\right]}\right\}. 
\end{equation}

\noindent Now only the first 6 matrices

\begin{equation}
\label{eq46}
\left\{\Gamma_{2\mathrm{N}}^{\underline{\mathrm{A}}}\right\}=\left\{\Gamma_{2\mathrm{N}}^{1},\Gamma_{2\mathrm{N}}^{2},\Gamma_{2\mathrm{N}}^{3},\Gamma_{2\mathrm{N}}^{4},\Gamma_{2\mathrm{N}}^{5},\Gamma_{2\mathrm{N}}^{6}\right\} 
\end{equation}

\noindent from the set (38) play the role of generating operators in the constructions (45), (46).

The maximal pure matrix algebra of invariance of the FW equation (29) is the 32 dimensional algebra SO(6)$\oplus i\Gamma_{2\mathrm{N}}^{0}\cdot$SO(6)$\oplus i\Gamma_{2\mathrm{N}}^{0}$.

The additional possibilities, which are open by the 29 orts of the algebra SO(8) in comparison with 16 orts of the well-known algebra SO(1,5), are principal in description of the Bose states in the framework of the Dirac theory [69--73]. The algebra SO(8) includes two independent spin s=1/2 SU(2) subalgebras $(s^{1}\equiv \frac{1}{2}\Gamma_{2\mathrm{N}}^{2}\Gamma_{2\mathrm{N}}^{3}, \, s^{2}\equiv \frac{1}{2}\Gamma_{2\mathrm{N}}^{3}\Gamma_{2\mathrm{N}}^{1}, \, s^{3}\equiv \frac{1}{2}\Gamma_{2\mathrm{N}}^{1}\Gamma_{2\mathrm{N}}^{2})$ and ($\check{s}^{1}\equiv \frac{1}{2}\Gamma_{2\mathrm{N}}^{5}\Gamma_{2\mathrm{N}}^{6}, \, \check{s}^{2}\equiv \frac{1}{2}\Gamma_{2\mathrm{N}}^{6}\Gamma_{2\mathrm{N}}^{4}, \, \check{s}^{3}\equiv \frac{1}{2}\Gamma_{2\mathrm{N}}^{4}\Gamma_{2\mathrm{N}}^{5}$) (in the anti-Hermitian form), whereas the SO(1,5) algebra includes only one set of SU(2) generators given by the elements ($s^{1}\equiv \frac{1}{2}\Gamma_{2\mathrm{N}}^{2}\Gamma_{2\mathrm{N}}^{3}, \, s^{2}\equiv \frac{1}{2}\Gamma_{2\mathrm{N}}^{3}\Gamma_{2\mathrm{N}}^{1}, \, s^{3}\equiv \frac{1}{2}\Gamma_{2\mathrm{N}}^{1}\Gamma_{2\mathrm{N}}^{2}$). Moreover, the algebra SO(6), which is the algebra of invariance of the FW equation (29), includes these two independent SU(2) subalgebras as well. As long as these two spin s=1/2 SU(2) sets of generators commute between each other, their combination gives the generators of spin s=1 representation of SU(2) algebra. Such SU(2) algebra is the building element in construction of spin s=1 Lorentz and Poincar$\mathrm{\acute{e}}$ algebras, with respect to which the FW equation (29) is invariant. 

The transition to the RCQM is given by the transformation (25), (26). Thus, in the quantum-mechanical representation the 7 $\Gamma$ matrices (38) (in the terms of standard $\Gamma_{2\mathrm{N}}^{\bar{\mu}}$ matrices (29), (30)) have the form (33) together with 

\begin{equation}
\label{eq47} \bar{\Gamma}_{2\mathrm{N}}^{5} \equiv \Gamma_{2\mathrm{N}}^{1}\Gamma_{2\mathrm{N}}^{3}\widehat{C}, \quad \bar{\Gamma}_{2\mathrm{N}}^{6} \equiv -i\Gamma_{2\mathrm{N}}^{2}\Gamma_{2\mathrm{N}}^{4}\widehat{C}, \quad \bar{\Gamma}_{2\mathrm{N}}^{7} \equiv i; \quad \bar{\gamma}^{1}\bar{\gamma}^{2}\bar{\gamma}^{3}\bar{\gamma}^{6}\bar{\gamma}^{5}\bar{\gamma}^{6}\bar{\gamma}^{7}=\mathrm{I}_{4},
\end{equation}

\noindent and satisfy the anticommutation relations of the Clifford--Dirac algebra Cl$^{\texttt{R}}$(0,6) representation in the following form

\begin{equation}
\label{eq48} \bar{\Gamma}_{2\mathrm{N}}^{\mathrm{A}} \bar{\Gamma}_{2\mathrm{N}}^{\mathrm{B}} + \bar{\Gamma}_{2\mathrm{N}}^{\mathrm{B}}\bar{\Gamma}_{2\mathrm{N}}^{\mathrm{A}} = -
2\delta^{\mathrm{A}\mathrm{B}},\quad \mathrm{A},\mathrm{B}=\overline{1,7}.
\end{equation}

The RCQM representation of the algebra SO(8) is given by

\begin{equation}
\label{eq49}
\bar{s}^{\bar{\mathrm{A}}\bar{\mathrm{B}}}=\{\bar{s}^{\mathrm{A}\mathrm{B}}=\frac{1}{4}[\bar{\Gamma}_{2\mathrm{N}}^{\mathrm{A}},\bar{\Gamma}_{2\mathrm{N}}^{\mathrm{B}}], \, \bar{s}^{\mathrm{A}8}=-\bar{s}^{8\mathrm{A}}=\frac{1}{2}\bar{\Gamma}_{2\mathrm{N}}^{\mathrm{A}}\},\quad \bar{\mathrm{A}},\bar{\mathrm{B}}=\overline{1,8},
\end{equation}

\noindent where the matrices $\bar{\Gamma}_{2\mathrm{N}}^{\mathrm{A}}$ are given in (33), (47) and generators $\bar{s}^{\bar{\mathrm{A}}\bar{\mathrm{B}}}$ satisfy the commutation relations

\begin{equation}
\label{eq50}
[\bar{s}^{\bar{\mathrm{A}}\bar{\mathrm{B}}},\bar{s}^{\bar{\mathrm{C}}\bar{\mathrm{D}}}]=
\delta^{\bar{\mathrm{A}}\bar{\mathrm{C}}}\bar{s}^{\bar{\mathrm{B}}\bar{\mathrm{D}}}
+\delta^{\bar{\mathrm{C}}\bar{\mathrm{B}}}\bar{s}^{\bar{\mathrm{D}}\bar{\mathrm{A}}}
+\delta^{\bar{\mathrm{B}}\bar{\mathrm{D}}}\bar{s}^{\bar{\mathrm{A}}\bar{\mathrm{C}}}
+\delta^{\bar{\mathrm{D}}\bar{\mathrm{A}}}\bar{s}^{\bar{\mathrm{C}}\bar{\mathrm{B}}}.
\end{equation}

The RCQM representation of the algebra SO(6)$\subset$SO(8) is given by

\begin{equation}
\label{eq51}
\left\{\mathrm{I}, \, \alpha^{\underline{\mathrm{A}} \,\underline{\mathrm{B}}} =
2\bar{s}^{\underline{\mathrm{A}} \, \underline{\mathrm{B}}}\right\},  \quad  \underline{\mathrm{A}},\underline{\mathrm{B}} = \overline{1,6},
\end{equation}

\noindent where

\begin{equation}
\label{eq52}
\left\{\bar{s}^{\underline{\mathrm{A}} \,\underline{\mathrm{B}}}\right\}=\left\{\bar{s}^{\underline{\mathrm{A}} \,\underline{\mathrm{B}}}\equiv \frac{1}{4}{\left[\bar{\Gamma}_{2\mathrm{N}}^{\underline{\mathrm{A}}},\bar{\Gamma}_{2\mathrm{N}}^{\underline{\mathrm{B}}}\right]}\right\}. 
\end{equation}

The maximal pure matrix algebra of invariance of the Schr$\mathrm{\ddot{o}}$dinger--Foldy equation (13) is the 32 dimensional algebra $\mathrm{a}_{32}=$SO(6)$\oplus i\cdot$SO(6)$\oplus i$.

In the RCQM representation two independent sets of generators of SU(2) subalgebra of SO(6) algebra, with respect to which the Schr$\mathrm{\ddot{o}}$dinger--Foldy equation (13) is invariant, have the form 

\begin{equation}
\label{eq53}
\bar{s}^{1}\equiv \frac{1}{2}\bar{\Gamma}_{2\mathrm{N}}^{2}\bar{\Gamma}_{2\mathrm{N}}^{3}, \, \bar{s}^{2}\equiv \frac{1}{2}\bar{\Gamma}_{2\mathrm{N}}^{3}\bar{\Gamma}_{2\mathrm{N}}^{1}, \, \bar{s}^{3}\equiv \frac{1}{2}\bar{\Gamma}_{2\mathrm{N}}^{1}\bar{\Gamma}_{2\mathrm{N}}^{2}, 
\end{equation}

\begin{equation}
\label{eq54}
\bar{\check{s}}^{1}\equiv \frac{1}{2}\bar{\Gamma}_{2\mathrm{N}}^{5}\bar{\Gamma}_{2\mathrm{N}}^{6}, \, \bar{\check{s}}^{2}\equiv \frac{1}{2}\bar{\Gamma}_{2\mathrm{N}}^{6}\bar{\Gamma}_{2\mathrm{N}}^{4}, \, \bar{\check{s}}\equiv \frac{1}{2}\bar{\Gamma}_{2\mathrm{N}}^{4}\bar{\Gamma}_{2\mathrm{N}}^{5}, 
\end{equation}

\noindent where the matrices $\bar{\Gamma}_{2\mathrm{N}}^{\mathrm{A}}$ are given in (33), (47). The situation here is similar to the FW representation. As long as these two spin s=1/2 SU(2) sets of generators commute between each other, their combination gives the generators of spin s=1 representation of SU(2) algebra. Such SU(2) algebra is the building element in construction of spin s=1 Lorentz and Poincar$\mathrm{\acute{e}}$ algebras, with respect to which the Schr$\mathrm{\ddot{o}}$dinger--Foldy equation (13) is invariant.

The partial case of 4 component formalism and $4 \times 4$ $\gamma$ matrices algebra follows from the above given consideration after corresponding substitutions 2N=4, etc. The improved consideration of this axiom in [1] for the spin 1/2 particle-antiparticle doublet should be taken from the text above as the corresponding particular case. 

An important fact is that in Pauli--Dirac representation the 32 dimensional algebra $\mathrm{SO(6)}\oplus i\Gamma_{2\mathrm{N}}^{0}\cdot \mathrm{SO(6)}\oplus i\Gamma_{2\mathrm{N}}^{0}$ still is the algebra of invariance of the Dirac equation (or 2N component Dirac-like equation). The only difference is that in Pauli--Dirac representation the operators ${s}^{\underline{\mathrm{A}} \, \underline{\mathrm{B}}}$ and gamma matrices $\Gamma_{2\mathrm{N}}^{\underline{\mathrm{A}}}$ are not pure matrix. They are the nonlocal pseudo-differential operators, which contain the operator $\widehat{\omega}\equiv \sqrt{-\Delta+m^{2}}$. For the standard 4 component Dirac equation the corresponding gamma matrices are given by (18)--(22) in [73].

Another interesting fact is that operator of the Dirac equation with Hamiltonian $H \equiv \gamma^{0}\overrightarrow{\gamma} \cdot \overrightarrow{p}+\gamma^{0}m-e^{2}/\left|\overrightarrow{x}\right|$ commutes with all 32 generators of the $\mathrm{SO(6)}\oplus i\gamma^{0}\cdot \mathrm{SO(6)}\oplus i\gamma^{0}$ algebra, which elements are given in terms of gamma matrices from (18)--(22) in [73]. Therefore, the relativistic hydrogen atom has wide additional symmetries given by the 31 nontrivial generators from $\mathrm{SO(6)}\oplus i\gamma^{0}\cdot \mathrm{SO(6)}\oplus i\gamma^{0}$. Moreover, the relativistic hydrogen atom has additional spin 1 symmetries, which for the free Dirac equation are known from [69--73] and which are the consequences of the algebra $\mathrm{SO(6)}\oplus i\gamma^{0}\cdot \mathrm{SO(6)}\oplus i\gamma^{0}$.

Note that consideration of this axiom above is given as the simple generalization of the correct spin 1/2 particle-antiparticle doublet formalism. In reality the dimensions of Clifford--Dirac (and SO(n)) algebras for the higher spin cases 2(2s+1)$\geq$8 can be more higher then considered above 64 dimensional Cl$^{\texttt{R}}$(4,2) and Cl$^{\texttt{R}}$(0,6) algebras (and SO(8) algebra). The reason of this situation is the possibility to construct for 2(2s+1)$\geq$8 cases the additional $\Gamma$ matrices obeying the Clifford--Dirac anticommutation relations. The concrete examples of such additional $\Gamma$ matrices can be found in [81, 82].
 
\textbf{On the main and additional conservation laws}. Similarly to the nonrelativistic quantum mechanics \textit{the conservation laws are found in the form of quantum-mechanical mean values of the operators, which commute with the operator of the equation of motion}.

The important physical consequence of the assertion about the relativistic invariance is the fact that 10 integral dynamical variables of the doublet

\begin{equation}
\label{eq55}
(P_{\mu}, \, J_{\mu\nu}) \equiv \int d^{3}x f^{\dag}(t,\overrightarrow{x})(\widehat{p}_{\mu}, \, \widehat{j}_{\mu\nu})f(t,\overrightarrow{x})=\mathrm{Const}
\end{equation}

\noindent do not depend on time, i. e. they are the constants of motion for this doublet. In (55) and below in the text of this axiom the wave function is chosen as (4) and generators ($\widehat{p}_{\mu}, \, \widehat{j}_{\mu\nu}$) are given in (12), (13) of [1].

Note that the external and internal degrees of freedom  for the free arbitrary spin particle-antiparticle doublet are independent. Therefore, the operator $\overrightarrow{s}$ (15) commutes not only with the operators $\widehat{\overrightarrow{p}}, \overrightarrow{x}$, but also with the orbital part $\widehat{m}_{\mu\nu}$ of the total angular momentum operator. And both operators $\overrightarrow{s}$ and $\widehat{m}_{\mu\nu}$ commute with the operator $i\partial_{t}-\sqrt { - \Delta + m^2}$ of the equation (13). Therefore, besides the 10 main (consequences of the 10 Poincar$\mathrm{\acute{e}}$ generators) conservation laws (55), 12 additional constants of motion exist for the free arbitrary spin particle-antiparticle doublet. These additional conservation laws are the consequences of the operators of the following observables (index 2N is omitted):

\begin{equation}
\label{eq56}
s_{j}, \quad \breve{s}_{\ell}=\frac{s_{\ell n}\widehat{p}_{n}}{\widehat{\omega}+m}, \quad \widehat{m}_{\ell n}=x_{l}\widehat{p}_{n}-x_{n}\widehat{p}_{\ell}, \quad \widehat{m}_{0 \ell}=-\widehat{m}_{l0}=t\widehat{p}_{\ell}-\frac{1}{2}\left\{x_{\ell},\widehat{\omega}\right\},
\end{equation}

\noindent part of which are given and explained in (21). Here $s_{j}=s_{\ell n}$ are given in (15).

Thus, the following assertions can be proved. In the space $\mathrm{H}^{\mathrm{A}}=\left\{A\right\}$ of the quantum-mechanical amplitudes the 10 main conservation laws (55) have the form

\begin{equation}
\label{eq57}
(P_{\mu},J_{\mu\nu})=\int d^{3}k A^{\dag}(\overrightarrow{k})(\check{\widetilde{p}}_{\mu},\check{\widetilde{j}}_{\mu\nu})A(\overrightarrow{k}), \quad A(\overrightarrow{k})\equiv \left|
{{\begin{array}{*{20}c}
a^{1}(\overrightarrow{k})\\
a^{2}(\overrightarrow{k})\\
\cdot\\
\cdot\\
a^{2\mathrm{N}}(\overrightarrow{k})\\ 
\end{array} }} \right|,
\end{equation}

\noindent where the density generators of $\mathcal{P}^{\mathrm{A}}$, $(\check{\widetilde{p}}_{\mu},\check{\widetilde{j}}_{\mu\nu})$ from (57) are given by

\begin{equation}
\label{eq58}
\check{\widetilde{p}}_{0}=p_{0}=\omega, \, \check{\widetilde{p}}_{\ell}=p_{\ell}=k_{\ell}, \, \check{\widetilde{j}}_{\ell n}=\widetilde{x}_{\ell}k_{n}-\widetilde{x}_{n}k_{\ell}+s_{\ell n}; \quad (\widetilde{x}_{l}=-i\frac{\partial}{\partial k^{l}}),
\end{equation}
\begin{equation}
\label{eq59}
\check{\widetilde{j}}_{0\ell}=-\frac{1}{2}\left\{\widetilde{x}_{\ell},\omega \right\}-(\breve{\widetilde{s}}_{\ell}\equiv\frac{s_{\ell n}k_{n}}{\omega+m}).
\end{equation}

\noindent In the formula (57) $A(\overrightarrow{k})$ is a 2N-column of amplitudes.

Note that the operators (57)--(59) satisfy the Poincar$\mathrm{\acute{e}}$ commutation relations in the manifestly covariant form (11).

It is evident that the 12 additional conservation laws

\begin{equation}
\label{eq60}
(M_{\mu \nu}, \, S_{\ell n}, \, \breve{S}_{\ell}) \equiv\int d^{3}x f^{\dag}(t,\overrightarrow{x})(\widehat{m}_{\mu \nu}, \, s_{\ell n}, \, \breve{s}_{\ell})f(t,\overrightarrow{x})
\end{equation}

\noindent generated by the operators (56), are the separate terms in the expressions (57)--(59) of principal (main) conservation laws.

The 22 above given conservation laws are valid for an arbitrary spin (in the case s=0 we have only 10 conservation laws generated by operators (21)).

Additional set of conservation laws is given by 32 quantities 

\begin{equation}
\label{eq61}
(\mathrm{A}_{32}) \equiv \int d^{3}x f^{\dag}(t,\overrightarrow{x})(\widehat\mathrm{a}_{32})f(t,\overrightarrow{x})=\mathrm{Const},
\end{equation}

\noindent or in terms of quantum-mechanical amplitudes

\begin{equation}
\label{eq62}
(\mathrm{A}_{32})=\int d^{3}k A^{\dag}(\overrightarrow{k})(\widehat\mathrm{a}_{32})A(\overrightarrow{k}).
\end{equation}

\noindent Here ($\widehat\mathrm{a}_{32}$) are the pure matrix generators of the algebra $\mathrm{a}_{32}=$SO(6)$\oplus i\cdot$SO(6)$\oplus i$.

In the case of spin s=1/2 the conservation law of spin is given twice: in the set (60) and in the set (62). For cases s>1/2 the sets (60) and (62) are completely different. 

\textbf{On the stationary complete sets of operators}. Let us consider now the outstanding role of the different \textit{complete sets} of operators from the algebra of observables $\mathrm{A}_{\mathrm{S}}$. If one does not appeal to the complete sets of operators, then the solutions of the the Schr$\mathrm{\ddot{o}}$dinger-Foldy equation (13) are linked directly only with the Sturm-Liouville problem for the energy operator (5). In this case one comes to so-called "degeneration" of solutions. Recall that for an arbitrary complete sets of operators the notion of degeneration is absent in the Sturm-Liouville problem (see, e.g., [65]): only one state vector corresponds to any one point of the common spectrum of a complete set of operators. To wit,  for a complete set of operators there is a one to one correspondence between any point of the common spectrum and an eigenvector.

The \textit{stationary complete sets} play the special role among the complete sets of operators. Recall that the stationary complete set is the set of all functionally independent mutually commuting operators, each of which commutes with the operator of energy (in our case with the operator (5)). The examples of the stationary complete sets in $\mathrm{H}^{3,2\mathrm{N}}$ are given by $(\widehat{\overrightarrow{p}}, \, s_{z}\equiv s^{3}_{2\mathrm{N}}, \, g)$, $(\overrightarrow{p}, \, \overrightarrow{s}_{2\mathrm{N}}\cdot\overrightarrow{p}, \, g)$, ets; $g$ is the charge sign operator. The set $(\overrightarrow{x}, \, s_{z}, \, g)$ is an example of non-stationary complete set. The $\overrightarrow{x}$-realization (10) of the space $\mathrm{H}^{3,2\mathrm{N}}$ and of quantum-mechanical Schr$\mathrm{\ddot{o}}$dinger-Foldy equation (13) are related just to this complete set.

For the goals of this paper the stationary complete set $(\widehat{\overrightarrow{p}}, \, s_{z}\equiv s^{3}_{2\mathrm{N}}, \, g)$ is chosen. The series of partial examples of equations on eigenvectors and eigenvalues for this stationary complete set is given in [1]. Consider here as an example the spin s=1/2 particle-antiparticle doublet case. Corresponding equations on eigenvectors and eigenvalues are given by

\begin{equation}
\label{eq63} \widehat{\overrightarrow{p}}e^{-ikx}\mathrm{d}_{\alpha} =
\overrightarrow{k}e^{-ikx}\mathrm{d}_{\alpha}, \quad \alpha =1,2,3,4,
\end{equation}

\begin{equation}
\label{eq64}
s^{3}\mathrm{d}_{1} = \frac{1}{2}\mathrm{d}_{1}, \, s^{3}\mathrm{d}_{2} = -\frac{1}{2}\mathrm{d}_{2}, \, s^{3}\mathrm{d}_{3} = -\frac{1}{2} \mathrm{d}_{3}, \, s^{3}\mathrm{d}_{4} = \frac{1}{2}\mathrm{d}_{4},
\end{equation}
\vskip1mm

\begin{equation}
\label{eq65} g\mathrm{d}_{1} = -\mathrm{d}_{1}, \,
g\mathrm{d}_{2} = -\mathrm{d}_{2}, \,  g\mathrm{d}_{3} =
\mathrm{d}_{3}, \, g\mathrm{d}_{4} = \mathrm{d}_{4},
\end{equation}

\noindent where the Cartesian orts $\{\mathrm{d}_{\alpha}\}$ are
given in [1] in the formulae (42).

Comparison of the given in equations (64) spin s=1/2 case with the spin s=2 case ((194) in [1]) shows that the general form is out of clear visualization. Thus, the appealing to such general form is absent here. 

The interpretation of the amplitudes in the general solution (14) follows from equations of (63)--(65) type for the fixed value of arbitrary spin. In general, the functions $a^{2\mathrm {N}}(\overrightarrow{k})$ are the quantum-mechanical momentum-spin amplitudes. Thus, the first half of functions $a^{2\mathrm {N}}(\overrightarrow{k})$ is the momentum-spin amplitudes of the particle (e. g., electron) with the momentum $\widehat{\overrightarrow{p}}$, sign of the charge ($-e$) and corresponding spin projections (for the electron, e. g., $\frac{1}{2}, \, -\frac{1}{2}$), respectively. Further, the second (bottom) half of functions $a^{2\mathrm {N}}(\overrightarrow{k})$ is the momentum-spin amplitudes of the antiparticle (e. g., positron) with the momentum $\widehat{\overrightarrow{p}}$, sign of the charge ($+e$) and opposite spin projections (for the positron, e. g.,$-\frac{1}{2}, \, \frac{1}{2}$), respectively.

Thus, the conclusion about the fermionic (or bosonic) spin features of the solution (14) (i. e. the interpretation of the solution (14)) follows from the equations on eigenvectors and eigenvalues (of (63)--(65) type) and the above given interpretation of the amplitudes.

\textbf{On the solutions of the Schr$\mathrm{\ddot{o}}$dinger-Foldy equation}. Let us consider the Schr$\mathrm{\ddot{o}}$dinger-Foldy equation (13) general solution  related to the stationary complete sets $(\widehat{\overrightarrow{p}}, \, s_{z}\equiv s^{3}, \, g)$, where $s^{3}$ is given in (15). The fundamental solutions of the equation (13), which are the eigen solutions of this stationary complete sets, are given by the relativistic de Broglie waves:

\begin{equation}
\label{eq66}
\varphi_{\vec{k}2\mathrm {N}} (t,\overrightarrow{x})=\frac{1}{(2\pi)^{\frac{3}{2}}}e^{-i\omega t + i\vec{k}\vec{x} }\mathrm{d}_{2\mathrm {N}},
\end{equation}
\vskip3mm

\noindent where $\mathrm{d}_{2\mathrm {N}}$ are the orts of 2N dimensional Cartesian basis (the Cartesian orts are the common eigenvectors for the operators $(s_{z}, \, g)$).

Vectors (66) are the generalized solutions of the equation (13). These solutions do not belong to the quantum-mechanical space $\mathrm{H}^{3,2\mathrm {N}}$, i. e. they are not realized in the nature. Nevertheless, the solutions (66) are the complete orthonormalized orts in the rigged Hilbert space $\mathrm{S}^{3,2\mathrm {N}}\subset\mathrm{H}^{3,2\mathrm {N}}\subset\mathrm{S}^{3,2\mathrm {N}*}$. In symbolic form the conditions of orthonormalisation and completeness are given by

\begin{equation}
\label{eq67}
\int d^{3}x\varphi^{\dag}_{\vec{k}\alpha}(t,\overrightarrow{x})\varphi_{\vec{k}^{\prime}\alpha^{\prime}}(t,\overrightarrow{x})=\delta(\overrightarrow{k}-\overrightarrow{k}^{\prime})\delta_{\alpha\alpha^{\prime}},
\end{equation}
\begin{equation}
\label{eq68}
\int d^{3}k\sum_{\alpha=1}^{2\mathrm {N}}\varphi^{\beta}_{\vec{k}\alpha}(t,\overrightarrow{x})\varphi^{*\beta^{\prime}}_{\vec{k}\alpha}(t,\overrightarrow{x}^{\prime})=\delta(\overrightarrow{x}-\overrightarrow{x}^{\prime})\delta_{\beta\beta^{\prime}}.
\end{equation}

\noindent The functional forms of these conditions are omitted because of their bulkiness.

In the rigged Hilbert space $\mathrm{S}^{3,2\mathrm {N}}\subset\mathrm{H}^{3,2\mathrm {N}}\subset\mathrm{S}^{3,2\mathrm {N}*}$ an arbitrary solution of the equation (13) can be decomposed in terms of fundamental solutions (66). Furthermore, for the solutions $f\in\mathrm{S}^{3,2\mathrm {N}}\subset\mathrm{H}^{3,2\mathrm {N}}$ the expansion (14) is, (i) mathematically well-defined in the framework of the standard differential and integral calculus, (ii) if in the expansion (14) a state $f\in\mathrm{S}^{3,2\mathrm {N}}\subset\mathrm{H}^{3,2\mathrm {N}}$, then the amplitudes $a^{2\mathrm {N}}(\overrightarrow{k})$ in (14) belong to the set of the Schwartz test functions over $\mathrm{R}^{3}_{\vec{k}}$. Therefore, they have the unambiguous physical sense of the amplitudes of probability distributions over the eigen values of the stationary complete sets $(\widehat{\overrightarrow{p}}, \, s_{z}, \, g)$. Moreover, the complete set of quantum-mechanical amplitudes unambiguously determine the corresponding representation of the space $\mathrm{H}^{3,2\mathrm {N}}$ (in this case it is the $(\overrightarrow{k}, \, s_{z}, \, g)$-representation), which vectors have the harmonic time dependence

\begin{equation}
\label{eq69}
\widetilde{f}(t,\overrightarrow{k})=e^{-i\omega t}A(\overrightarrow{k}), \quad A(\overrightarrow{k})\equiv \left|
{{\begin{array}{*{20}c}
a^{1}(\overrightarrow{k})\\
a^{2}(\overrightarrow{k})\\
\cdot\\
\cdot\\
a^{2\mathrm{N}}(\overrightarrow{k})\\ 
\end{array} }} \right|,
\end{equation}

\noindent i. e. are the states with the positive sign of the energy $\omega$.

The similar assertion is valid for the expansions of the states $f\in\mathrm{H}^{3,2\mathrm {N}}$ over the basis states, which are the eigenvectors of an arbitrary stationary complete sets. Therefore, the corresponding representation of the space $\mathrm{H}^{3,2\mathrm {N}}$, which is related to such expansions, is often called as the generalized Fourier transformation.

By the way, the $\overrightarrow{x}$-realization (10) of the states space is associated with the non-stationary complete set of operators $(\overrightarrow{x}, \, s_{z}, \, g)$. Therefore, the amplitudes $f^{2\mathrm {N}}(t,\overrightarrow{x})=\mathrm{d}_{2\mathrm {N}}^{\dag}f(t,\overrightarrow{x})=U(t)f(0,\overrightarrow{x})$ of the probability distribution over the eigen values of this complete set depend on time $t$ non-harmonically.

\textbf{The axiom on the mean value of the operators of observables}. Note that any apparatus can not fulfill the absolutely precise measurement of a value of the physical quantity having continuous spectrum. Therefore, the customary quantum-mechanical axiom about the possibility of "precise" measurement, for example, of the coordinate (or another quantity with the continuous spectrum), which is usually associated with the corresponding "reduction" of the wave-packet, can be revisited. This assertion for the values with the continuous spectrum can be replaced by the axiom that only the mean value of the operator of observable (or the corresponding complete set of observables) is the experimentally observed for $\forall f\in\mathrm{H}^{3,\mathrm {N}}$. Such axiom, without any loss of generality of consideration, unambiguously justifies the using of the subspace $\mathrm{S}^{3,\mathrm {N}}\subset\mathrm{H}^{3,\mathrm {N}}$ as an approximative space of the physically realizable states of the considered object. This axiom as well does not enforce the application of the conception of the ray in $\mathrm{H}^{3,\mathrm {N}}$ (the set of the vectors $e^{i\alpha}f$ with an arbitrary-fixed real number $\alpha$) as the state of the object. Therefore, the mapping $(a, \, \varpi)\rightarrow U(a, \, \varpi)$ in the formula (22) for the $\mathcal{P}$-representations in $\mathrm{S}^{3,\mathrm {N}}\subset\mathrm{H}^{3,\mathrm {N}}$ is an unambiguous. Such axiom actually removes the problem of the wave packet "reduction", which discussion started from the well-known von Neumann monograph [36]. Therefore, the subjects of the discussions of all "paradoxes" of quantum mechanics, a lot of attention to which was paid in the past century, are removed also.

The important conclusion about the RCQM is as follows. The consideration of all aspects of this model is given on the basis of using only such conceptions and quantities, which have the direct relation to the experimentally observable physical quantities of this "elementary" physical system.

\textbf{On the principles of heredity and the correspondence}. The explicit forms (55)--(60) of the main and additional conservation laws demonstrate evidently that the model of RCQM satisfies the principles of the heredity and the correspondence with the non-relativistic classical and quantum theories. The deep analogy between RCQM and these theories for the physical system with the finite number degrees of freedom (where the values of the free dynamical conserved quantities are additive) is also evident.

Our new way of the Dirac equation derivation (section 9 in [1]) has been started from the RCQM of the spin s=(1/2,1/2) particle-antiparticle doublet.

\textbf{On the second quantization}. This axiom is external (not internal) in the RCQM. It is necessary for quantum field theory. The formalism of RCQM is complete without this axiom. The reader can see the brief consideration in [1] (section 8) in the example of spin s=(1/2,1/2) particle-antiparticle doublet.

\textbf{On the physical interpretation}. The physical interpretation always is the final step in the arbitrary model of the physical reality formulation. In the beginning of the interpretation is better to recall the different formulations of the quantum theory collected, e. g., in [83].  

The above considered model of physical reality is called the canonical quantum mechanics due to the principles of the heredity and correspondence with nonrelativistic Schr$\mathrm{\ddot{o}}$dinger quantum mechanics [36].

The above considered canonical quantum mechanics is called the relativistic canonical quantum mechanics due to its invariance with respect to the corresponding representations of the Poincar$\mathrm{\acute{e}}$ group $\mathcal{P}$.

The above considered RCQM describes the spin (s,s) particle-antiparticle doublet due to the corresponding eigenvalues in the equations like (63)--(65) for the stationary complete set of operators and the explicit forms of the Casimir operators ((15), (16) in [1]) of the corresponding Poincar$\mathrm{\acute{e}}$ group $\mathcal{P}$ representation, with respect to which the dynamical equation of motion is invariant.

The axioms of this section  eventually need to be reconciled with three levels of description used in this paper: RCQM, canonical FW and Dirac models. Nevertheless, this interesting problem cannot be considered in few pages.  The readers of this paper can compare the axioms of RCQM given above with the main principles of the Dirac model given in B. Thaller's monograph [84] on the high mathematical level.

\vskip 0.3cm

\begin{center}
\textbf{Section 7. General description of the arbitrary spin field theory}
\end{center}

\vskip 0.3cm

The step by step consideration of the different partial examples in [1] (sections 21--27) enabled us to rewrite them in the general form, which is valid for arbitrary spin. Therefore, the generalization of the consideration given in [1] leads to the general formalism of the arbitrary spin fields.

The formalism presented below in this section is valid for an arbitrary particle-antiparticle multiplet in general and for the particle-antiparticle doublet in particular.  

\textbf{The canonical (FW type) model of the arbitrary spin particle-antiparticle field}.

The operator, which transform the RCQM of the arbitrary spin particle-antiparticle multiplet into the corresponding canonical particle-antiparticle field, is given by (25). As it is explained already in section 7 above (axiom on the Clifford--Dirac algebra) the transition with the help of the operator (25) is possible for the anti-Hermitian operators. 

The formulas mentioned below are found from the corresponding formulas of RCQM with the help of the operator (25) on the basis of its properties (26), (27). For the general form of arbitrary spin canonical particle-antiparticle field the equation of motion of the FW type is given by (29). The general solution has the form (28), where $a^{\mathrm{N}}(\overrightarrow{k})$ are the quantum-mechanical momentum-spin amplitudes of the particle and $a^{\breve{\mathrm{N}}}(\overrightarrow{k})$ are the quantum-mechanical momentum-spin amplitudes of the antiparticle, $\left\{\mathrm{d}\right\}$ is 2N-component Cartesian basis.

The spin operator, which follows from (15), has the form

\begin{equation}
\label{eq70} \overrightarrow{s}_{2\mathrm{N}}= \left| {{\begin{array}{*{20}c}
 \overrightarrow{s}_{\mathrm{N}} \hfill  & 0 \hfill \\
 0 \hfill & \overrightarrow{s}_{\mathrm{N}} \hfill \\
 \end{array} }} \right|, \quad \mathrm{N}=2s+1,
\end{equation}
\vskip3mm

\noindent where $\overrightarrow{s}_{\mathrm{N}}$ are $\mathrm{N} \times \mathrm{N}$ generators of arbitrary spin irreducible representations of SU(2) algebra, which satisfy the commutation relations (18).

The generators of the reducible unitary representation of the
Poincar$\mathrm{\acute{e}}$ group $\mathcal{P}$, with respect to which the canonical field equation (29) and the set $\left\{\phi\right\}$ of its solutions (28) are invariant, are
given by

\begin{equation}
\label{eq71} \widehat{p}^{0}=\Gamma_{2\mathrm{N}}^{0}\widehat{\omega}\equiv \Gamma_{2\mathrm{N}}^{0}\sqrt{-\Delta+m^{2}}, \quad \widehat{p}^{\ell}=-i\partial_{\ell}, \quad \widehat{j}^{\ell
n}=x^{\ell}\widehat{p}^{n}-x^{n}\widehat{p}^{\ell}+s^{\ell
n}_{2\mathrm{N}} \equiv \widehat{m}^{\ell n}+s^{\ell n}_{2\mathrm{N}},
\end{equation}
\begin{equation}
\label{eq72} \widehat{j}^{0 \ell}=-\widehat{j}^{\ell
0}=x^{0}\widehat{p}^{\ell}-\frac{1}{2}\Gamma_{2\mathrm{N}}^{0}\left\{x^{\ell},\widehat{\omega}\right\}+\Gamma_{2\mathrm{N}}^{0}\frac{(\overrightarrow{s}_{2\mathrm{N}} \times
\overrightarrow{p})^{\ell}}{\widehat{\omega}+m},
\end{equation}
\vskip3mm

\noindent where arbitrary spin SU(2) generators
$\overrightarrow{s}_{2\mathrm{N}}=(s^{\ell n}_{2\mathrm{N}})$ have the form (70), $\Gamma_{2\mathrm{N}}^{0}$ is given in (29).

Note that together with the generators (71), (72) another set of 10 operators commutes with the operator of equation (29), satisfies the commutation relations (11) of the Lie algebra of 
Poincar$\mathrm{\acute{e}}$ group $\mathcal{P}$, and, therefore, can be chosen as the Poincar$\mathrm{\acute{e}}$ symmetry of the model under consideration. This second set is given by the generators $\widehat{p}^{0}, \, \widehat{p}^{\ell}$ from (71) together with the orbital parts of the generators $\widehat{j}^{\ell
n}, \, \widehat{j}^{0 \ell}$ from (71), (72), respectively.

The calculation of the Casimir operators $p^{2}=\widehat{p}^{\mu}\widehat{p}_{\mu}, \, W=w^{\mu}w_{\mu}$ ($w^{\mu}$ is the Pauli--Lubanski pseudovector) for the fixed value of spin completes the brief description of the model.

\textbf{The locally covariant model of the arbitrary spin particle-antiparticle field}.

The operator, which transform the canonical (FW type) model of the arbitrary spin particle-antiparticle field into the corresponding locally covariant particle-antiparticle field, is the generalized FW operator and is given by

\begin{equation}
\label{eq73} V^{\mp}=\frac{\mp
\overrightarrow{\Gamma}_{2\mathrm{N}} \cdot\overrightarrow{p}+\widehat{\omega}+m}{\sqrt{2\widehat{\omega}(\widehat{\omega}+m)}},
\quad V^{-}=(V^{+})^{\dag}, \quad V^{-}V^{+}=V^{+}V^{-}=\mathrm{I}_{2\mathrm{N}}, \quad \mathrm{N}=2s+1,
\end{equation}

\noindent where $\Gamma_{2\mathrm{N}}^{j}$ are known from (30) and $\Sigma_{\mathrm{N}}^{j}$ are the $\mathrm{N} \times \mathrm{N}$ Pauli matrices. 

Of course, for the matrices $\Gamma_{2\mathrm{N}}^{\mu}$ (29), (30) (together with the matrix $\Gamma_{2\mathrm{N}}^{4}\equiv \Gamma_{2\mathrm{N}}^{0}\Gamma_{2\mathrm{N}}^{1}\Gamma_{2\mathrm{N}}^{2}\Gamma_{2\mathrm{N}}^{3}$) the relations (24) are valid.

Note that in formulas (73) and in all formulas before the end of the section the values of N are only even. Therefore, the canonical field equation (29) describes the larger number of multiplets then the generalized Dirac equation (74) given below. 

The formulas (74)--(79) below are found from the corresponding formulas (28), (29), (70)--(72) of canonical field model on the basis of the operator (73).

For the general form of arbitrary spin locally covariant particle-antiparticle field the Dirac-like equation of motion follows from the equation (29) after the transformation (73) and is given by

\begin{equation}
\label{eq74}
\left[i\partial_{0}-\Gamma_{2\mathrm{N}}^{0}(\overrightarrow{\Gamma}_{2\mathrm{N}}\cdot\overrightarrow{p}+m)\right]\psi(x)=0.
\end{equation}
\vskip3mm

The general solution has the form

\begin{equation}
\label{eq75}
\psi(x)=V^{-}\phi(x)= \frac{1}{\left(2\pi\right)^{\frac{3}{2}}}\int d^{3}k\left[e^{-ikx}a^{\mathrm{N}}(\overrightarrow{k})\mathrm{v}^{-}_{\mathrm{N}}(\overrightarrow{k})+e^{ikx}a^{*\breve{\mathrm{N}}}(\overrightarrow{k})\mathrm{v}^{+}_{\breve{\mathrm{N}}}(\overrightarrow{k})\right],
\end{equation}

\noindent where amplitudes and notation $\breve{\mathrm{N}}$ are the same as in (28); $\left\{\mathrm{v}^{-}_{\mathrm{N}}(\overrightarrow{k}), \, \mathrm{v}^{+}_{\breve{\mathrm{N}}}(\overrightarrow{k}) \right\}$ are 2N-component Dirac basis spinors with properties of orthonormalisation and completeness similar to 4-component Dirac spinors from [85]. 

The spin operator is given by

\begin{equation}
\label{eq76}
\overrightarrow{s}_{\mathrm{D}} =
V^{-}\overrightarrow{s}_{2\mathrm{N}}V^{+},
\end{equation}
\vskip3mm

\noindent where operator $\overrightarrow{s}_{2\mathrm{N}}$ is known from (70). The explicit forms of few partial cases of spin operators (76) are given in formulae (259)--(261), (284)--(286), (359) of [1] for the particle-antiparticle multiplets s=(1,0,1,0), s=(3/2,3/2), s=(2,1,2,1), respectively.

The generators of the reducible unitary representation of the
Poincar$\mathrm{\acute{e}}$ group $\mathcal{P}$, with respect to which the covariant field equation (74) and the set $\left\{\psi\right\}$ of its solutions (75) are invariant, have the form

\begin{equation}
\label{eq77} \widehat{p}^{0}=\Gamma_{2\mathrm{N}}^{0}(\overrightarrow{\Gamma}_{2\mathrm{N}}\cdot\overrightarrow{p}+m), \quad \widehat{p}^{\ell}=-i\partial_{\ell}, \quad \widehat{j}^{\ell
n}=x_{\mathrm{D}}^{\ell}\widehat{p}^{n}-x_{\mathrm{D}}^{n}\widehat{p}^{\ell}+s_{\mathrm{D}}^{\ell
n}\equiv \widehat{m}^{\ell n}+s_{\mathrm{D}}^{\ell n},
\end{equation}
\begin{equation}
\label{eq78} \widehat{j}^{0 \ell}=-\widehat{j}^{\ell
0}=x^{0}\widehat{p}^{\ell}-\frac{1}{2}\left\{x_{\mathrm{D}}^{\ell},\widehat{p}^{0}\right\}+\frac{\widehat{p}^{0}(\overrightarrow{s}_{\mathrm{D}}\times
\overrightarrow{p})^{\ell}}{\widehat{\omega}(\widehat{\omega}+m)},
\end{equation}

\noindent where the spin matrices
$\overrightarrow{s}_{\mathrm{D}}=(s_{\mathrm{D}}^{\ell n})$ are
given in (76) and the operator
$\overrightarrow{x}_{\mathrm{D}}$ has the form

\begin{equation}
\label{eq79}
\overrightarrow{x}_{\mathrm{D}}=\overrightarrow{x}+\frac{i\overrightarrow{\Gamma}_{2\mathrm{N}}}{2\widehat{\omega}}-\frac{\overrightarrow{s}_{2\mathrm{N}}^{\Gamma}\times \overrightarrow{p}}{\widehat{\omega}(\widehat{\omega}+m)}-\frac{i\overrightarrow{p} (\overrightarrow{\Gamma}_{2\mathrm{N}}\cdot \overrightarrow{p})}{2\widehat{\omega}^{2}(\widehat{\omega}+m)},
\end{equation}

\noindent where specific spin matrices $\overrightarrow{s}_{2\mathrm{N}}^{\Gamma}$ are given by

\begin{equation}
\label{eq80}
\overrightarrow{s}_{2\mathrm{N}}^{\Gamma} \equiv \overrightarrow{s}_{\mathrm{FW}} =\left(s^{1}_{2\mathrm{N}},s^{2}_{2\mathrm{N}},s^{3}_{2\mathrm{N}}\right)=\frac{i}{2}(\Gamma_{2\mathrm{N}}^{2}\Gamma_{2\mathrm{N}}^{3}, \, \Gamma_{2\mathrm{N}}^{3}\Gamma_{2\mathrm{N}}^{1}, \, \Gamma_{2\mathrm{N}}^{1}\Gamma_{2\mathrm{N}}^{2}).
\end{equation}

Note that for corresponding partial cases of $\overrightarrow{x}_{\mathrm{D}}$ in [1] ((267) for s=(1,0,1,0), once more (267) but for s=(3/2,3/2), (333) for s=(2,0,2,0), (363) for s=(2,1,2,1)) the corresponding $\overrightarrow{s}_{2\mathrm{N}}^{\Gamma}$ are given by

\begin{equation}
\label{eq81}
\mathrm{s}=(1,0,1,0) \, \mathrm{and} \,\, \mathrm{s}=(3/2,3/2): \quad \overrightarrow{s}_{8}^{\Gamma} \equiv \left(s^{1}_{8},s^{2}_{8},s^{3}_{8}\right)=\frac{i}{2}(\Gamma_{8}^{2}\Gamma_{8}^{3}, \, \Gamma_{8}^{3}\Gamma_{8}^{1}, \, \Gamma_{8}^{1}\Gamma_{8}^{2}),
\end{equation}

\noindent for $\overrightarrow{x}_{\mathrm{D}}$ [1] (267), where the $\Gamma_{8}^{j}$ matrices are given in (253),

\begin{equation}
\label{eq82}
\mathrm{s}=(2,0,2,0): \quad \overrightarrow{s}_{12}^{\Gamma} \equiv \left(s^{1}_{12},s^{2}_{12},s^{3}_{12}\right)=\frac{i}{2}(\Gamma_{12}^{2}\Gamma_{12}^{3}, \, \Gamma_{12}^{3}\Gamma_{12}^{1}, \, \Gamma_{12}^{1}\Gamma_{12}^{2}),
\end{equation}

\noindent for $\overrightarrow{x}_{\mathrm{D}}$ [1] (333), where the $\Gamma_{12}^{j}$ matrices are given in (324),

\begin{equation}
\label{eq83}
\mathrm{s}=(2,1,2,1): \quad \overrightarrow{s}_{16}^{\Gamma} \equiv \left(s^{1}_{16},s^{2}_{16},s^{3}_{16}\right)=\frac{i}{2}(\Gamma_{16}^{2}\Gamma_{16}^{3}, \, \Gamma_{16}^{3}\Gamma_{16}^{1}, \, \Gamma_{16}^{1}\Gamma_{16}^{2}),
\end{equation}

\noindent for $\overrightarrow{x}_{\mathrm{D}}$ [1] (363), where the $\Gamma_{16}^{j}$ matrices are given in (353),

It is easy to verify that the generators (77), (78) for any N commute with the operator of equation (74), and satisfy the commutation relations ((11) of [1]) of the Lie algebra of the Poincar$\mathrm{\acute{e}}$ group. The last step in the brief description of the model is the calculation of the Casimir operators $p^{2}=\widehat{p}^{\mu}\widehat{p}_{\mu}, \, W=w^{\mu}w_{\mu}$ ($w^{\mu}$ is the Pauli--Lubanski pseudovector) for the fixed value of spin.

\textbf{The example of spin s=(0,0) particle-antiparticle doublet}.

The completeness of simplest spin multiplets and doublets consideration of [1] is achieved by the supplementation of this example. The formalism follows from the general formalism of arbitrary spin after the substitution s=0.

The Schr$\mathrm{\ddot{o}}$dinger--Foldy equation of RCQM is given by (13) for N=1, i. e. it is 2-component equation. The solution is given by (14) for N=1. The Poincar$\mathrm{\acute{e}}$ group $\mathcal{P}$ generators, with respect to which the equation (13) for s=(0,0) is invariant, are given by (12), (13) of [1] taken in the form of $2 \times 2$ matrices with spin terms equal to zero, i. e. the corresponding generators are given by (21). 

The corresponding FW type equation of canonical field theory is given by

\begin{equation}
\label{eq84} (i\partial_{0}-
\sigma^{3}\widehat{\omega})\phi(x)=0, \quad \sigma^{3}=\left|
{{\begin{array}{*{20}c}
 1 \hfill & 0 \hfill \\
 0 \hfill & { -1} \hfill \\
\end{array} }} \right|, \quad \widehat{\omega}\equiv \sqrt{-\Delta+m^{2}}.
\end{equation}

The general solution is given by

\begin{equation}
\label{eq85} \phi(x)=
\frac{1}{\left(2\pi\right)^{\frac{3}{2}}}\int
d^{3}k\left[e^{-ikx}a^{1}(\overrightarrow{k})\mathrm{d}_{1}+e^{ikx}a^{*2}(\overrightarrow{k})\mathrm{d}_{2}\right].
\end{equation}
\vskip3mm

The Poincar$\mathrm{\acute{e}}$ group $\mathcal{P}$ generators, with respect to which the equation (84) and the set $\left\{\phi\right\}$  of its solutions (85) are invariant, have the form

\begin{equation}
\label{eq86} \widehat{p}^{0}=\sigma^{3}\widehat{\omega}\equiv \sigma^{3}\sqrt{-\Delta+m^{2}}, \quad \widehat{p}^{\ell}=-i\partial_{\ell}, \quad \widehat{j}^{\ell
n}=x^{\ell}\widehat{p}^{n}-x^{n}\widehat{p}^{\ell},
\end{equation}
\begin{equation}
\label{eq87} \widehat{j}^{0 \ell}=-\widehat{j}^{\ell
0}=x^{0}\widehat{p}^{\ell}-\frac{1}{2}\sigma^{3}\left\{x^{\ell},\widehat{\omega}\right\}.
\end{equation}
\vskip3mm

\noindent Generators (86), (87) are the partial $2 \times 2$ matrix form of operators (71), (72) taken with the spin terms equal to zero.

\textbf{Additional consideration of covariant field equation for spin s=(3/2,3/2) particle-antiparticle doublet}.

Consider the nontrivial partial example of covariant field equation for arbitrary spin. Such example is given by covariant field equation for spin s=(3/2,3/2) particle-antiparticle doublet. This case presents the demonstrative example how new equations can be derived by the developed in [1] and here methods.

Now contrary to [1] equation for spin s=3/2 is found as the simple partial case of general equation (74):

\begin{equation}
\label{eq88}
\left[i\partial_{0}-\Gamma_{8}^{0}(\overrightarrow{\Gamma}_{8}\cdot\overrightarrow{p}+m)\right]\psi(x)=0.
\end{equation}

\noindent Here the $\Gamma_{8}^{\mu}$ matrices are given by

\begin{equation}
\label{eq89} \Gamma_{8}^{0}=\left| {{\begin{array}{*{20}c}
 \mathrm{I}_{4} \hfill & 0 \\
 0 \hfill & -\mathrm{I}_{4} \\
\end{array} }} \right|, \quad \Gamma_{8}^{j}=\left| {{\begin{array}{*{20}c}
 0 \hfill & \Sigma^{j} \\
 -\Sigma^{j} \hfill & 0 \\
\end{array} }} \right|,
\end{equation}

\noindent where $\Sigma^{j}$ are the $4\times 4$ Pauli matrices

\begin{equation}
\label{eq90} \Sigma^{j}=\left| {{\begin{array}{*{20}c}
 \sigma^{j} \hfill & 0 \\
 0 \hfill & \sigma^{j} \\
\end{array} }} \right|,
\end{equation}

\noindent and $\sigma^{j}$ are the standard $2\times 2$ Pauli matrices. The matrices $\Sigma^{j}$ satisfy the similar commutation relations as the standard $2\times 2$ Pauli matrices and have other similar properties. The matrices $\Gamma_{8}^{\mu}$ (89) satisfy the anticommutation relations of the Clifford--Dirac algebra in the form (24) with N=4.

Note that equation (88) is not the ordinary direct sum
of the two Dirac equations. Therefore, it is not the complex
Dirac--Kahler equation [82]. Moreover, it is not the standard 16 component
Dirac--Kahler equation [86].

The solution of equation (88) is derived as a partial case from the solution (75) of
the general equation (74) and is given by
\begin{equation}
\label{eq91}
\psi(x)=V^{-}_{8}\phi(x)= \frac{1}{\left(2\pi\right)^{\frac{3}{2}}}\int d^{3}k\left[e^{-ikx}b^{\mathrm{A}}(\overrightarrow{k})\mathrm{v}^{-}_{\mathrm{A}}(\overrightarrow{k})+e^{ikx}b^{*\mathrm{B}}(\overrightarrow{k})\mathrm{v}^{+}_{\mathrm{B}}(\overrightarrow{k})\right],
\end{equation}

\noindent where $\mathrm{A}=\overline{1,4}, \, \mathrm{B}=\overline{5,8}$ and the 8-component spinors $(\mathrm{v}^{-}_{\mathrm{A}}(\overrightarrow{k}), \, \mathrm{v}^{+}_{\mathrm{B}}(\overrightarrow{k}))$ are given by (257) in [1].

The spinors $(\mathrm{v}^{-}_{\mathrm{A}}(\overrightarrow{k}), \, \mathrm{v}^{+}_{\mathrm{B}}(\overrightarrow{k}))$ satisfy the relations of the orthonormalization and completeness similar to the corresponding relations for the standard 4-component Dirac spinors, see, e. g., [85]. 

In the covariant local field theory, the operators of the SU(2) spin,
which satisfy the corresponding commutation relations $
\left[s_{8\mathrm{D}}^{j},s_{8\mathrm{D}}^{\ell}\right]=i\varepsilon^{j
\ell n}s_{8\mathrm{D}}^{n}$ and commute with the operator
$\left[i\partial_{0}-\Gamma_{8}^{0}(\overrightarrow{\Gamma}_{8}\cdot\overrightarrow{p}+m)\right]$
of equation (88), are derived from the pure matrix operators
(279) of [1] with the help of transition operator $V^{-}_{8}$: $\overrightarrow{s}_{8\mathrm{D}}=V^{-}_{8}\overrightarrow{s}_{8}V^{+}_{8}$. The explicit form of the transition operator $V^{\mp}_{8}$ is given in (249)--(251) of [1]. The explicit form of these s=(3/2,3/2) SU(2) generators was given already by formulae (284)--(287) in [1].

The equations on eigenvectors and eigenvalues of the operator
$s_{8\mathrm{D}}^{3}$ (286) in [1] follow from the equations (280) of [1] and
the transformation $V^{-}_{8}$. In addition to it, the action of
the operator $s_{8\mathrm{D}}^{3}$ (286) in [1] on the spinors
$(\mathrm{v}^{-}_{\mathrm{A}}(\overrightarrow{k}), \,
\mathrm{v}^{+}_{\mathrm{B}}(\overrightarrow{k}))$ (257) in [1] also
leads to the result
\vskip3mm
\noindent $s_{8\mathrm{D}}^{3}\mathrm{v}^{-}_{1}(\overrightarrow{k})= \frac{3}{2}\mathrm{v}^{-}_{1}(\overrightarrow{k}), \, s_{8\mathrm{D}}^{3}\mathrm{v}^{-}_{2}(\overrightarrow{k}) = \frac{1}{2}\mathrm{v}^{-}_{2}(\overrightarrow{k}), \, s_{8\mathrm{D}}^{3}\mathrm{v}^{-}_{3}(\overrightarrow{k}) = -\frac{1}{2}\mathrm{v}^{-}_{3}(\overrightarrow{k}), \,
s_{8\mathrm{D}}^{3}\mathrm{v}^{-}_{4}(\overrightarrow{k}) = -\frac{3}{2}\mathrm{v}^{-}_{4}(\overrightarrow{k}),$

\begin{equation}
\label{eq92}
s_{8\mathrm{D}}^{3}\mathrm{v}^{+}_{5}(\overrightarrow{k}) =
\frac{3}{2}\mathrm{v}^{+}_{5}(\overrightarrow{k}), \,
s_{8\mathrm{D}}^{3}\mathrm{v}^{+}_{6}(\overrightarrow{k}) = \frac{1}{2}\mathrm{v}^{+}_{6}(\overrightarrow{k}), \, s_{8\mathrm{D}}^{3}\mathrm{v}^{+}_{7}(\overrightarrow{k}) = -\frac{1}{2}\mathrm{v}^{+}_{7}(\overrightarrow{k}), \, s_{8\mathrm{D}}^{3}\mathrm{v}^{+}_{8}(\overrightarrow{k}) = -\frac{3}{2}\mathrm{v}^{+}_{8}(\overrightarrow{k}).
\end{equation}

In order to verify equations (92) the identity $(\tilde{\omega}+m)^{2}+(\overrightarrow{k})^{2}=2\tilde{\omega}(\tilde{\omega}+m)$ is used. In the case $\mathrm{v}^{+}_{\mathrm{B}}(\overrightarrow{k})$ in the
expression $s_{8\mathrm{D}}^{3}(\overrightarrow{k})$ (286) of [1] the substitution $\overrightarrow{k}\rightarrow - \overrightarrow{k}$ is made.

The equations (92) determine the interpretation of the amplitudes in solution (91). Nevertheless, the direct quantum-mechanical interpretation of the amplitudes should be made in the framework of the RCQM and is already given in [1] (section 14 in paragraph after equations (183)).

The explicit form of the $\mathcal{P}$-generators of the fermionic
representation of the Poincar$\mathrm{\acute{e}}$ group
$\mathcal{P}$, with respect to which the covariant equation (88)
and the set $\left\{\psi\right\}$ of its solutions (91) are
invariant, is derived as a partial case from the generators (77), (78). The corresponding
generators are given by 

\begin{equation}
\label{eq93} \widehat{p}^{0}=\Gamma_{8}^{0}(\overrightarrow{\Gamma}_{8}\cdot\overrightarrow{p}+m), \quad \widehat{p}^{\ell}=-i\partial_{\ell}, \quad \widehat{j}^{\ell
n}=x_{\mathrm{D}}^{\ell}\widehat{p}^{n}-x_{\mathrm{D}}^{n}\widehat{p}^{\ell}+s_{8\mathrm{D}}^{\ell
n}\equiv \widehat{m}^{\ell n}+s_{8\mathrm{D}}^{\ell n},
\end{equation}
\begin{equation}
\label{eq94} \widehat{j}^{0 \ell}=-\widehat{j}^{\ell
0}=x^{0}\widehat{p}^{\ell}-\frac{1}{2}\left\{x_{\mathrm{D}}^{\ell},\widehat{p}^{0}\right\}+\frac{\widehat{p}^{0}(\overrightarrow{s}_{8\mathrm{D}}\times
\overrightarrow{p})^{\ell}}{\widehat{\omega}(\widehat{\omega}+m)},
\end{equation}

\noindent where the spin matrices $\overrightarrow{s}_{8\mathrm{D}}=(s_{8\mathrm{D}}^{\ell n})$ are
given by (284)--(286) in [1] and the operator $\overrightarrow{x}_{\mathrm{D}}$ has the form

\begin{equation}
\label{eq95}
\overrightarrow{x}_{\mathrm{D}}=\overrightarrow{x}+\frac{i\overrightarrow{\Gamma}_{8}}{2\widehat{\omega}}-\frac{\overrightarrow{s}_{8}^{\Gamma}\times \overrightarrow{p}}{\widehat{\omega}(\widehat{\omega}+m)}-\frac{i\overrightarrow{p} (\overrightarrow{\Gamma}_{8}\cdot \overrightarrow{p})}{2\widehat{\omega}^{2}(\widehat{\omega}+m)},
\end{equation}

\noindent with specific spin matrices $\overrightarrow{s}_{8}^{\Gamma}$ given in (81).

It is easy to verify that the generators (93), (94) with SU(2) spin (284)--(286) from [1] commute with
the operator $\left[i\partial_{0}-\Gamma_{8}^{0}(\overrightarrow{\Gamma}_{8}\cdot\overrightarrow{p}+m)\right]$
of equation (88), satisfy the commutation relations ((11) of [1]) of the
Lie algebra of the Poincar$\mathrm{\acute{e}}$ group and the corresponding Casimir operators are given by $p^{2}=\widehat{p}^{\mu}\widehat{p}_{\mu}=m^{2}\mathrm{I}_{8}, \, W=w^{\mu}w_{\mu}=m^{2}\overrightarrow{s}_{8\mathrm{D}}^{2}=\frac{3}{2}\left(\frac{3}{2}+1\right)m^{2}\mathrm{I}_{8}$.

The conclusion that equation (88) describes the local field of fermionic
particle-antiparticle doublet of the spin s=(3/2,3/2) and mass
$m>0$ (and its solution (91) is the local fermionic field of the above
mentioned spin and nonzero mass) follows from the analysis of
equations (92) and the above given calculation of the Casimir operators $p^{2}, \, W=w^{\mu}w_{\mu}$.

Hence, the equation (88) describes the spin s=(3/2,3/2) particle-antiparticle doublet on the same level, on which the standard 4-component Dirac equation describes the spin s=(1/2,1/2) particle-antiparticle doublet. Moreover, the external argument in the validity of such interpretation is the link with the corresponding RCQM of spin s=(3/2,3/2) particle-antiparticle doublet, where the quantum-mechanical interpretation is direct and evident. Therefore, the fermionic spin s=(3/2,3/2) properties of equation (88) are proved.

Contrary to the bosonic spin s=(1,0,1,0) properties of the equation  (88) found in [1] (section 22), the fermionic spin s=(1/2,1/2,1/2,1/2) properties of this equation are evident. The fact that equation (88) describes the multiplet of two fermions with the spin s=1/2 and two antifermions with that spin can be proved much more easier then the above given consideration. The proof is similar to that given in the standard 4-component Dirac model. The detailed consideration can be found in sections 7, 9, 10 of [1]. Therefore, equation (88) has more extended property of the Fermi--Bose duality then the standard Dirac equation [69--73]. This equation has the property of the Fermi--Bose triality. The property of the Fermi--Bose triality of the manifestly covariant equation (88) means that this equation describes on equal level (i) the spin s=(1/2,1/2,1/2,1/2) multiplet of two spin s=(1/2,1/2) fermions and two spin s=(1/2,1/2) antifermions, (ii) the spin s=(1,0,1,0) multiplet of the vector and scalar bosons together with their antiparticles, (iii) the spin s=(3/2,3/2) particle-antiparticle doublet.

It is evident that equation (88) is new in comparison with the Pauli--Fierz [47], Rarita--Schwinger [48] and Davydov [49] equations for the spin s=3/2 particle. Contrary to 16-component equations from [47--49] equation (88) is 8-component and does not need any additional condition. Formally equation (88) looks like to have some similar features with the Bargman--Wigner equation [62] for arbitrary spin, when the spin value is taken 3/2. The transformation $V^{\mp}_{8}=\frac{\mp
\overrightarrow{\Gamma}_{8}\cdot\overrightarrow{p}+\widehat{\omega}+m}{\sqrt{2\widehat{\omega}(\widehat{\omega}+m)}}$ looks like the transformation of Pursey [63] in the case of s=3/2. Nevertheless, the difference is clear. The given here model is derived from the first principles of RCQM (not from the FW type representation of the canonical field theory). Our consideration is original and new. The link with corresponding RCQM, the proof of the symmetry properties, the well defined spin operator (284)--(286) in [1], the features of the Fermi--Bose duality (triality) of the equation (88), the interaction with electromagnetic field and many other characteristics are suggested firstly.

Interaction, quantization and Lagrange approach in the above given spin s=(3/2,3/2) model are completely similar to the Dirac 4-component theory and standard quantum electrodynamics. For example, the Lagrange function of the system of interacting 8 component spinor and electromagnetic fields (in the terms of 4-vector potential $A^{\mu}(x)$) is given by

\begin{equation}
\label{eq96} L=-\frac{1}{4}F^{\mu\nu}F_{\mu\nu}+ \frac{i}{2} \left(\overline{\psi}(x)\Gamma_{8}^{\mu}\frac{\partial \psi (x)}{\partial x^{\mu}}-\frac{\partial \overline{\psi}(x)}{\partial x^{\mu}}\Gamma_{8}^{\mu}\psi (x)\right)-m\overline{\psi}(x)\psi (x)+q\overline{\psi}(x)\Gamma_{8}^{\mu}\psi (x)A_{\mu}(x),
\end{equation}

\noindent where $\overline{\psi}(x)$ is the independent Lagrange variable and $\overline{\psi}=\psi^{\dag}\Gamma_{8}^{0}$ in the space of solutions $\left\{\psi\right\}$. In Lagrangian (96) $F_{\mu\nu}=\partial _{\mu}A_{\nu}-\partial _{\nu}A_{\mu}$ is the electromagnetic field tensor in the terms of potentials, which play the role of variational variables in this Lagrange approach.

Therefore, the covariant local quantum field theory model for the interacting particles with spin s=3/2 and photons can be constructed in complete analogy to the construction of the modern quantum electrodynamics. This model can be useful for the investigations of processes with interacting hyperons and photons.

\vskip 0.3cm

\begin{center}
\textbf{Section 8. Briefly on the  different ways of the Dirac equation derivation}
\end{center}

\vskip 0.3cm

Among the results of this paper the original method of the derivation of the standard Dirac and the Dirac-like equation for arbitrary spin is suggested. In order to determine the place of this derivation among the other known methods consider below the different ways of the Dirac equation derivation. 

One should note the elegant derivation given by Paul Dirac in his book [3]. Until today it is very interesting for the readers to feel Dirac's way of thinking and to follow his logical steps. Nevertheless, the Dirac's consideration of the Schr$\mathrm{\ddot{o}}$dinger--Foldy equation, which was essentially used in his derivation [3], was not correct. Especially his assertion that Schr$\mathrm{\ddot{o}}$dinger--Foldy equation is unsatisfactory from the point of view of the relativistic theory. Dirac's point of view is considered here in section 2. Dirac's doubts were overcome in [4, 5, 12]. Today the significance of the Schr$\mathrm{\ddot{o}}$dinger--Foldy equation is confirmed by more then a hundred publications about FW (see, e. g., [29, 44, 63, 87] and the references therein) and the spinless Salpeter equations [6, 11, 22--28, 30, 31, 33, 34], which have wide-range application in contemporary theoretical physics.  

In the well-known book [85] one can find a review of the Dirac theory and two different ways of the
Dirac equation derivation. First, it is the presentation of the
Klein--Gordon equation in the form of a first-order
differential system of equations, factorization of the
Klein--Gordon operator. Second, the Lagrange approach is considered and the Dirac equation is derived
from the variational Euler--Lagrange least action principle.

In van der Waerden--Sakurai derivation [88] of the Dirac equation the spin of the electron is incorporated into the nonrelativistic theory. The representation of the nonrelativistic kinetic energy operator of the free spin 1/2 particle in the form $H^{\mathrm{KE}} = (\overrightarrow{\sigma} \cdot \overrightarrow{p})(\overrightarrow{\sigma} \cdot \overrightarrow{p})/2m$ and the relativistic expression $E^{2}-\overrightarrow{p}^{2}=m^{2}$ are used. Then the procedure of transition from 2-component to 4-component equation is fulfilled and explained.

In the book [89] (second edition) the Dirac equation is
derived from the manifestly covariant transformational properties
of the 4-component spinor.

The derivation of the Dirac equation from the initial
geometric properties of the space-time and electron together with
wide-range discussion of the geometric principles of the electron
theory is the main content of the book [90]. The ideas of V. Fock
and D. Iwanenko [91, 92] on the geometrical sense of Dirac
$\gamma$-matrices are the basis of the approach.

One should point out the derivation of the Dirac equation
based on the Bargman--Wigner classification of the irreducible unitary
representations of the Poincar$\mathrm{\acute{e}}$ group, see e.
g. [93]. It is an illustrative demonstration of the possibilities
of the group-theoretical approach to the elementary particle
physics.

In L. Foldy's papers [4, 5, 12] one can easy find the inverse problem, in which the Dirac equation is obtained from the FW equation. Nevertheless, it is only the transition from one representation of the spinor field to another.

H. Sallhofer [94, 95] derived the Dirac equation for hydrogen spectrum starting from the Maxwell equations in medium. Strictly speaking, only the stationary equations were considered.

In [96] quaternion measurable processes were introduced and the Dirac equation was derived from the Langevin equation associated with a two-valued process.

The author of [97] was able to derive the Dirac equation from the conservation law of spin 1/2 current. The requirement that this current is conserved leads to a unique determination of the Lorentz invariant equation satisfied by the relativistic spin 1/2 field. Let us briefly comment that the complete list of conservation laws for the Dirac theory is the Noether consequence of the Dirac equation. Therefore, the validity of the inverse problem is really expected. Can it be considered an independent derivation?  

The Dirac equation was derived [98] from the master equation of Poisson process by analytic continuation. The extension to the case where a particle moves in an external field was given. It was shown that the generalized master equation is closely related to the three-dimensional Dirac equation in an external field.

In [99], a method of deriving the Dirac equation from the relativistic Newton's second law was suggested. Such derivation is possible in a new formalism, which relates the special form of relativistic mechanics to the quantum mechanics. The author suggested a concept of a velocity field. At first, the relativistic Newton's second law was rewritten as a field equation in terms of the velocity field, which directly reveals a new relationship linked to the quantum mechanics. After that it was shown that the Dirac equation can be derived from the field equation in a rigorous and consistent manner.

A geometrical derivation of the Dirac equation, by considering a spin 1/2 particle traveling
with the speed of light in a cubic spacetime lattice, was made in [100]. The mass of the particle acts to flip the multi-component wave function at the lattice sites. Starting with a difference equation for the
case of one spatial and one time dimensions, the authors generalize the approach to higher dimensions.
Interactions with external electromagnetic and gravitational fields are also considered. Nevertheless, the idea of such derivation is based on the Dirac's observation that the instantaneous velocity operators of the spin 1/2 particle (hereafter called by the generic name <<the electron>>) have eigenvalues $\pm c$. This mistake of Dirac was demonstrated and overcome in Ref. [4].

Using the mathematical tool of Hamilton's bi-quaternions, the authors of [101] propose a derivation of the Dirac equation from the geodesic equation. Such derivation is given in the program of application of the theory of scale relativity to the purposes of microphysics at recovering quantum mechanics as a new non-classical mechanics on a non-derivable space-time.

M. Evans was successful to express his equation of general relativity (generally covariant field equation for gravitation and electromagnetism [102]) in spinor form, thus producing the Dirac equation in general relativity [103]. The Dirac equation in special relativity is recovered in the limit of Euclidean or flat spacetime.

More then ten years ago we already presented our own derivation of the Dirac equation [104--106]. The Dirac equation was derived from slightly generalized Maxwell equations with gradient-like current and charge densities. This form of the Maxwell equations, which is directly linked with the Dirac equation, is the maximally symmetrical variant of these equations. Such Maxwell equations are invariant with respect to a 256-dimensional algebra (the well-known algebra of conformal group has only 15 generators). Of course, we derived only massless Dirac equation.

In our recent papers [9, 10], and in [1], the Dirac equation has been derived from the 4-component Schr$\mathrm{\ddot{o}}$dinger--Foldy equation (13) of the RCQM. The starting point RCQM model of the spin s=(1/2,1/2) particle-antiparticle doublet has been formulated in these articles as well. Hence, the Dirac equation has been derived from the more fundamental model of the same physical reality, which is presented by the RCQM of the spin s=(1/2,1/2) particle-antiparticle doublet.

In order to have a complete picture of the different ways of the Dirac equation derivations the references to the articles [107--109] should be given. 

In [107] the derivation of [89] was repeated together with taking into account the physical meaning of the negative energies and the relative intrinsic parity of the elementary particles. The authors believe that the derivation from [89] is improved. 

Author of [108] first determines that each eigenfunction of a bound particle is a specific superposition of plane wave states that fulfills the averaged energy relation. After that the Schrodinger and Dirac equations were derived as the unique conditions the wavefunction must satisfy at each point in order to fulfill the corresponding energy equation. The Dirac equation involving electromagnetic potentials has been derived.

It has been shown recently in [109], without using the relativity principle, how the Dirac equation in three space-dimensions emerges from the large-scale dynamics of the minimal nontrivial quantum cellular automaton satisfying unitarity, locality, homogeneity, and discrete isotropy. The Dirac equation is recovered for
small wave-vector and inertial mass, whereas Lorentz covariance is distorted in the ultra-relativistic
limit.

Thus, a review of the different derivations of the Dirac equation demonstrates that presented in [1] and here method of the Dirac equation (and the Dirac-like equation for arbitrary spin) derivation is original and new.

It follows from the above given consideration that new methods of the Dirac equation derivation are not stopped and \textit{to be continued}.

\vskip 0.3cm

\begin{center}
\textbf{Section 9. Interaction}
\end{center}

\vskip 0.3cm

This paper in general is about free non-interacting fields and particle states. Note at first that the free non-interacting fields and particle states are the physical reality of the same level as the interacting fields and corresponding particle states. Nevertheless, it is meant that the interaction between a fields can be easily introduced on the every step of consideration. One test model with interaction is considered in explicit form in the section 7 above. The interaction can not be the deficiency in these constructions and can be introduced in many places by the method similar to the formula (96).

Note at first that the free non-interacting fields and particle states are the physical reality of the same level as the interacting fields and corresponding particle states. The best well-known example is the free Maxwell electromagnetic field in the form of free electromagnetic waves. Therefore, the systematic investigation of free non-interacting fields and particle states on the different levels of the relativistic canonical quantum mechanics of arbitrary spin (based on the Schrodinger-Foldy equation), the relativistic canonical field theory of arbitrary spin (based on the FW equation and its generalizations), the locally covariant field theory of arbitrary spin (based on the Dirac equation and its generalizations) and, moreover, the introduction of the operator link between these theories is some result as well and, I hope, have some independent significance. The precise free field consideration is the necessary first step in all consequent field theory presentation, in axiomatic presentation as well. Further, it is well-known that when the free field models are well-defined the interaction between them can be introduced very easy in well-defined forms known from the literature. 

The free particle RCQM today is the subject of other authors considerations as well. Note that the wave packet solutions of the Salpeter equation and their unusual properties (rather recent papers [6, 32]) have been derived for the free non-interacting cases.

Moreover, for the standard Dirac theory the interaction is given in all handbooks on the relativistic quantum field theory and quantum electrodynamics. Such consideration can be added here very easy. Nevertheless, here one typical interaction is considered already as the test example in order to demonstrate the way of introduction of the interaction in corresponding models. For the spin s=(3/2,3/2) field the interaction with electromagnetic potentials was given in the section 7 in the formula (96). Therefore, I had a thought that it is enough to explain the situation with interaction here. The interaction can not be the deficiency in these constructions and can be introduced in many places by the method similar to the formula (96).

Other concrete physical effects and the problem of description of (inter)-particle interactions in frames of the RCQM are described by other authors, see the references [22--32, 33, 34].

\vskip 0.3cm

\begin{center}
\textbf{Section 10. Application to the discussion around the antiparticle negative mass}
\end{center}

\vskip 0.3cm

The system of vertical and horizontal links between the RCQM and the field theory, which is proved above, has different useful applications. One of the fundamental applications is the participation in the discussion around the antiparticle negative mass. We emphasize that the model of the RCQM and the corresponding field theory do not need the appealing to the antiparticle negative mass concept [110--113]. It is natural due to the following reasons.

It is only the energy which depends on the mass. And the total energy together with the momentum is related to the external degrees of freedom, which are common and the same for the particle and antiparticle (for the electron and positron). The difference between the particle and the antiparticle consists only in internal degrees of freedom such as the spin $\overrightarrow{s}$ and the sign of the charge $g=-\gamma^{0}$. Thus, if in the RCQM the mass of the particle is taken positive then the mass of the antiparticle must be taken positive too.

On the other hand, a comprehensive analysis [111] of the Dirac equation for the doublet had led the authors of Ref. [111] to the concept of the negative mass of the antiparticle. Therefore, the consideration here gives the additional arguments that the Dirac model (or the Foldy-Wouthuysen model related to it) is not the quantum-mechanical one. Furthermore, in the problem of the relativistic hydrogen atom the use of the negative-frequency part $\psi^{-}(x)=e^{-i\omega t}\psi(\overrightarrow{x})$ of the spinor $\psi(x)$ in the "role of the quantum-mechanical object" is not valid. In this case neither $\left|\psi(\overrightarrow{x})\right|^{2}$, nor $\overline{\psi}(\overrightarrow{x})\psi(\overrightarrow{x})$ is the probability distribution density with respect to the eigenvalues of the Fermi-doublet coordinate operator. It is due to the fact [4] that in the Dirac model the $\overrightarrow{x}$ is not the experimentally observable Fermi-doublet coordinate operator.

The application of the RCQM can be useful for the analysis of the experimental situation found in Ref. [114]. Such analysis is interesting due to the fact that (as it is demonstrated here and in [1]) the RCQM is the most fundamental model of the particle-antiparticle doublet.

Another interesting application of the RCQM is inspired by Ref. [115], where the quantum electrodynamics is reformulated in the FW representation. The author of Ref. [115] essentially used the result of Ref. [111] on the negative mass of the antiparticle. Starting from the RCQM we are able not to appeal to the concept of the antiparticle negative mass.

\vskip 0.3cm

\begin{center}
\textbf{Section 11. Discussions and conclusions}
\end{center}

\vskip 0.3cm

In the presented above text our experience [9, 10, 69--73, 104--106] in the time span 2002--2013 in the investigation of the spin s=1/2 and s=1 fields is applied for the first time to the higher spin cases s=3/2 and s=2 (in the form of electronic preprint it has been fulfilled for the first time in [1]). Thus, our "old" papers are augmented by the list of new results for higher spins and generalization for arbitrary spin. Moreover, here (the start has been given in [1]) the system of different vertical and horizontal links between the arbitrary spin particles descriptions on the levels of relativistic quantum mechanics, canonical field theory (of FW type) and locally covariant field theory is suggested.

Among the results of this paper the original method of derivation of the Dirac (and the Dirac-like equations for higher spins) is suggested (section 7). In order to determine the place of this derivation among the other known methods in section 8 the different ways of the Dirac equation derivation [3, 88--103, 107--109] have been reviewed. Thus, a review of the different derivations of the Dirac equation demonstrates that presented in section 7 general method is original and new. Here the Dirac equation is derived from the Schr$\mathrm{\ddot{o}}$dinger--Foldy equation (13) (in 4-component case) of the RCQM. The RCQM model of the spin s=(1/2,1/2) particle-antiparticle doublet has been considered in details in section 7 of [1]. Hence, the Dirac equation is derived in [1] and here from the more fundamental model of the same physical reality, which is presented by the RCQM of the spin s=(1/2,1/2) particle-antiparticle doublet.

One of the general fundamental conclusions is as follows. It is shown by corresponding comparison that customary FW representation can not give the complete quantum-mechanical description of the relativistic particle (or particle multiplet). Compare, e. g. the equations on eigenvectors and eigenvalues for the third component of the spin operator in each quantum-mechanical and canonical field theory model here above and in [1]. It is useful also to compare the general solutions in RCQM and in field theory (it is enough to consider the field theory in FW representation). Contrary to RCQM in FW representation the general solution consists of positive and negative frequency parts. As a consequence contrary to RCQM in FW representation the energy has an indefinite sign. Hence, the complete quantum-mechanical description of the relativistic particle (or particle multiplet) can be given only in the framework of the RCQM. Therefore, the customary FW transformation is extended here to the form, which gives the link between the locally covariant field theory and the RCQM. Hence, such extended inverse FW transformation is used here to fulfill the synthesis of covariant particle equations. The start of such synthesis is given here from the RCQM and not from the canonical field theory in the representations of the Foldy-Wouthuysen type.

Comparison of RCQM and FW representation visualizes the role of the J. von Neumann axiomatic [36] in this presentation. Therefore, relation of J. von Neumann axiomatic [36] to the overall contents of the paper is direct and unambiguous. It is shown that among the above considered models only RCQM of arbitrary spin can be formulated in J. von Neumann's axiomatic, whereas canonical and covariant field theories can not be formulated in its framework.

The new operator links $v_{2\mathrm{N}} = \left| {{\begin{array}{*{20}c}
 \mathrm{I}_{\mathrm{N}} \hfill &  0 \hfill\\
 0 \hfill & \hat{C}\mathrm{I}_{\mathrm{N}}  \hfill\\
 \end{array} }} \right|, \, \mathrm{N}=2s+1,$ found here between the RCQM of arbitrary spin and the canonical (FW type) field theory enabled ones to translate the result found in these models from one model to another. For example, the results of [25--28] from RCQM can be traslated into the canonical field theory. Contrary, the results of [29] from canonical field theory can be translated into the RCQM (for free non-interacting cases and in the form of anti-Hermitian operators). Note that operator (25) is not unitary but is well defined and having inverse operator. 

The partial case of the Schrodinger-Foldy equation, when the wave function has only one component [14], is called the spinless Salpeter equation and is under consideration in many recent papers [6, 27, 28, 30--35]. The partial wave packet solutions of this equation are given in [6]. The solutions for free massless and massive particle on a line, massless particle in a linear potential, plane wave solution for a free particle (these solution is given here in (8) for N-component case), free massless particle in three dimensions have been considered. Further, in the paper [32] other time dependent wave packet solutions of the free spinless Salpeter equation are given. Taking into account the relation of such wave packets to the Levy process the spinless Salpeter equation (in one dimensional space) is called in [32] as the Levy-Schrodinger equation. The several examples of the characteristic behavior of such wave packets have been shown, in particular, of the multimodality arising in their evolutions: a feature at variance with the typical diffusive unimodality of both the corresponding Levy process densities and usual Schrodinger wave functions. Therefore, the interesting task is to extend such consideration to the equations of the N-component relativistic canonical quantum mechanics considered above and to use the links given here in order to transform wave packet solutions [6, 32] into the solutions of the equations of the locally covariant field theory.

In this article the original FW transformation [4] is used and slightly generalized for the many component cases. The improvement of the FW transformation [4] is the task of many authors from 1950 until today, see, e. g., the recent publications [116--119]. Nevertheless, this transformation for free case of non-interacting spin 1/2 particle-antiparticle doublet is not changed from 1950 (the year of first publication) until today. Alexander Silenko was successful in FW transformation for single particles with spin 0 [116], spin 1/2 [117] and spin 1 [118, 119] interacting with external electric, magnetic and other fields. In the case of non-interacting particle, when the external electric, magnetic and other external fields are equal to zero, all results of [116--119] and other authors reduce to the earlier results [4, 120]. Therefore, the choice in this paper of the exact FW transformation from 1950 as the initial (and basic for further generalizations for arbitrary spin) is evident and well-defined. In our next articles, we will consider interacting fields and will use the results of [116--119] and recent results of other authors, which generalize the FW formulas in the case of interaction.

A few remarks should be added about the choice of the spin operator. Authors of recent paper [121] considered all spin operators for a Dirac particle satisfying some logical and group-theoretical conditions. The discussion of other spin operators proposed in the literature has been presented as well. As a result only one satisfactory operator has been chosen. This operator is equivalent to the Newton--Wigner spin operator and FW mean-spin operator. Contrary to such way the situation here is evident. Above the choice of the spin operator for spin s=1/2 particle-antiparticle doublet is unique. The explicit form for such operator follows directly from the main principles of the RCQM of spin s=1/2 particle-antiparticle doublet, which are formulated in the section 7 of [1]. Such operator from RCQM is given here in formula (34) taken for N=2. After that the links between the RCQM, FW representation and the Dirac model unambiguously give at first the FW spin (35) (with N=2) and finally the spin

\begin{equation}
\label{eq97}
\overrightarrow{s}_{\mathrm{D}}=\overrightarrow{s}-\frac{\overrightarrow{\gamma}\times
\nabla}{2\widehat{\omega}}+\frac{\nabla \times
(\overrightarrow{s}\times
\nabla)}{\widehat{\omega}(\widehat{\omega}+m)},
\end{equation}

\noindent which is the FW mean-spin operator (here $\overrightarrow{s}$ is given in (35) with N=2 and $\overrightarrow{\gamma}$ are $4 \times 4$ standard Dirac matrices). Therefore, the similar consideration for the higher spin doublets gives unambiguously the well-defined higher spin operators, which are presented in the sections 22--27 of [1]. These new mean-spin operators (259)--(261), (284)--(286), (359) in [1] for the N-component Dirac type equations for higher spins 1, 3/2 and 2 are the interesting independent results.

The goal of the paper [35] is a comprehensive analysis of the intimate relationship between jump-type stochastic processes (e. g. Levy flights) and nonlocal (due to integro-differential operators involved) quantum dynamics. Special attention is paid to the spinless Salpeter equation and the various wave packets, in particular to their radial expression in 3D. Furthermore, Foldy's approach [5] is used [35] to encompass free Maxwell theory, which however is devoid of any <<particle>> content. Links with the photon wave mechanics are explored. Consideration in [1], see e. g. the sections 13, 22 and 23, demonstrates another link between the Maxwell equations and the RCQM. In the generalization of the Foldy's synthesis of covariant particle equations given here (and in [1]) the Maxwell equations and their analogy for nonzero mass are related to the RCQM of spin s=(1,1) and spin s=(1,0,1,0) particle-antiparticle doublets, another approach see in [122, 123]. The electromagnetic field equations that follow from the corresponding relativistic quantum mechanical equations have been found in [1]. The new electrodynamical equations containing the hypothetical antiphoton and massless spinless antiboson have been introduced. The Maxwell-like equations for the boson with spin s=1 and $m>0$ (W-boson) have been introduced as well. In other words in [1], the Maxwell equations for the field with nonzero mass have been introduced.

The covariant consideration of arbitrary spin field theory given here and in sections 21--28 of [1] contains the non-covariant representations of the Poincar$\mathrm{\acute{e}}$ algebra. Nevertheless, it is not the deficiency of the given model. For the Poincar$\mathrm{\acute{e}}$ group $\mathcal{P}$ generators of spin s=(1/2,1/2) the covariant form 

\begin{equation}
\label{eq98} p^{\mu}= i\partial ^{\mu}, \quad j^{\mu
\nu}=x^{\mu}p^{\nu}-x^{\nu}p^{\mu}+s^{\mu \nu}, \quad s^{\mu \nu}\equiv \frac{i}{4}\left[\gamma^{\mu},\gamma^{\nu}\right], 
\end{equation}

\noindent is well-known, in which the generators have the form of the local Lie operators. Only in order to have the uniform consideration (77), (78) the Poincar$\mathrm{\acute{e}}$ generators for spin s=(1,0,1,0), (3/2,3/2), (2,0,2,0), (2,1,2,1) fields are given in formulae (265), (266), (331), (332), (361), (362) of [1] in corresponding uniform forms of non-covariant operators in covariant theory. Of course, the corrections given here in (80)--(83) for $\overrightarrow{s}_{2\mathrm{N}}^{\Gamma}$ are necessary. After further transformations of these generators sets in the direction of finding the covariant forms like (98) some sets of generators can be presented in the manifestly covariant forms. For other sets of generators covariant forms are extrinsic. Some sets of generators can be presented only in the forms, which are similar to given in [69--73], where the prime anti-Hermitian operators and specific eigenvectors -- eigenvalues equations (with imagine eigenvalues) are used, see, e. g., formula (21) in [70].

The second reason of the stop on the level (265), (266), (331), (332), (361), (362) of [1] is to conserve the important property of the Poincar$\mathrm{\acute{e}}$ generators in the canonical FW type representation. Similarly to the FW type Poincar$\mathrm{\acute{e}}$ generators in the sets (265), (266), (331), (332), (361), (362) of [1] both angular momenta (orbital and spin) commute with the operator of the Dirac-like equation of motion (74). Contrary to the generators (77), (78), in the covariant form (98) only total angular momentum, which is the sum of orbital and spin angular momenta, commutes with the Diracian.

The main point is as follows. The non-covariance is not the barrier for the relativistic invariance! Not a matter of fact that non-covariant objects such as the Lebesgue measure $d^{3}x$ and the non-covariant Poincar$\mathrm{\acute{e}}$ generators are explored, the model of locally covariant field theory of arbitrary spin presented in section 7 is a relativistic invariant in the following sense. The Dirac-like equation (74) and the set $\left\{\psi\right\}$ of its solutions (75) are invariant with respect to the reducible representation of the Poincar$\mathrm{\acute{e}}$ group $\mathcal{P}$ the non-local and non-covariant generators of which are given by (77), (78). Indeed, the direct calculations visualize that generators (77), (78) commute with the operator of equation (74) and satisfy the commutation relations ((11) in (1)) of the Lie algebra of the Poincar$\mathrm{\acute{e}}$ group $\mathcal{P}$.

The partial case of zero mass has been considered briefly in section 28 of [1].

The 8-component manifestly covariant equation (88) for the spin s=3/2 field found in [1] and here is the s=3/2 analogy of the 4-component Dirac equation for the spin s=1/2 doublet. It is shown that synthesis of this equation from the relativistic canonical quantum mechanics of the spin s=3/2 particle-antiparticle doublet is completely similar to the synthesis of the Dirac equation from the relativistic canonical quantum mechanics of the spin s=1/2 particle-antiparticle doublet. The difference is only in the value of spin (3/2 and 1/2). On this basis and on the basis of the investigation of solutions and transformation properties with respect to the Poincar$\mathrm{\acute{e}}$ group this new 8-component equation is suggested to be well defined for the description of spin s=3/2 fermions. Note that known Rarita-Schwinger (Pauli-Fierz) equation has 16 components and needs the additional condition.

The properties of the Fermi--Bose duality, triality and quadro Fermi--Bose properties of equations found have been discussed briefly.

Hence, the method of synthesis of manifestly covariant field equations on the basis of start from the relativistic canonical quantum mechanics of arbitrary spin is suggested. Its approbation on few principal examples is presented.

The main general conclusion is as follows. Among the three main models of arbitrary spin (relativistic canonical quantum mechanics, canonical field theory and covariant field theory) considered here the relativistic canonical quantum mechanics is the best in rigorous quantum-mechanical description. The transition from the relativistic canonical quantum mechanics to the canonical field theory essentially worsens the quantum-mechanical description. And the final transition from the canonical field theory to the covariant field theory essentially worsens the quantum-mechanical description once more.

Thus, the results of [1] are presented here in the general forms. After the given above small corrections all results of [1] are confirmed here.

The part of these results, which is related to the spin s=1, s=(1,1), s=(1,0,1,0) relativistic canonical quantum mechanics and electrodynamics, has been presented on 15-th MMET Conference [123]. Other examples of new equations have been considered recently on XXIIIth International Conference on Integrable Systems and Quantum symmetries [124] and on 8-th Symposium on Integrable Systems [125].

\textbf{Acknowledgments}
The author is much grateful for both unknown referees of the brief 22 pages journal version of this manuscript for very useful remarks and suggestions.

\begin{center}
\textbf{References}
\end{center}

\noindent 1. V.M. Simulik, "Covariant local field theory equations following from the relativistic canonical quantum mechanics of arbitrary spin," arXiv: 1409.2766v2 [quant-ph, hep-th] 19 Sep 2014, 67 p.

\vskip 0.3cm

\noindent 2. P.A.M. Dirac, "The quantum theory of the electron," \textit{Proc. R. Soc. Lond. A.}, vol. 117, no. 778, pp. 610--624, (1928).

\vskip 0.3cm

\noindent 3. P.A.M. Dirac, \textit{The principles of quantum mechanics, 4-th edition,} Oxford: Clarendon Press, 1958, 314 p.

\vskip 0.3cm

\noindent 4. L.L. Foldy and S.A. Wouthuysen, "On the Dirac theory of spin 1/2 particles and its non-relativistic limit," \textit{Phys. Rev.}, vol. 78, no. 1, pp. 29--36, (1950).

\vskip 0.3cm

\noindent 5. L.L. Foldy, "Synthesis of covariant particle equations," \textit{Phys. Rev.}, vol. 102, no. 2, pp. 568--581, (1956).

\vskip 0.3cm

\noindent 6. K. Kowalski and J. Rembielinski, "Salpeter equation and probability current in the relativistic Hamiltonian quantum mechanics," \textit{Phys. Rev. A.}, vol. 84, no. 1, pp. 012108 (1--11), (2011).

\vskip 0.3cm

\noindent 7. V. Simulik, I. Krivsky, and I. Lamer, "Lagrangian for the spinor field in the Foldy--Wouthuysen representation," \textit{Int. J. Theor. Math. Phys.}, vol. 3, no. 4, pp. 109--116, (2013). 

\vskip 0.3cm

\noindent 8. R.W. Brown, L.M. Krauss, and P.L. Taylor, "Leslie Lawrance Foldy," \textit{Physics Today}, vol. 54, no. 12, pp. 75--76, (2001).

\vskip 0.3cm

\noindent 9. V.M. Simulik and I.Yu. Krivsky, "Quantum-mechanical description of the fermionic doublet and its link with the Dirac equation," \textit{Ukr. J. Phys.}, vol. 58, no. 12, pp. 1192--1203, (2013).

\vskip 0.3cm

\noindent 10. V.M. Simulik and I.Yu. Krivsky, "Link between the relativistic canonical quantum mechanics and the Dirac equation," \textit{Univ. J. Phys. Appl.}, vol. 2, no. 2, pp. 115--128, (2014).

\vskip 0.3cm

\noindent 11. E.E. Salpeter, "Mass corrections to the fine structure of hydrogen-like atoms," \textit{Phys. Rev.}, vol. 87, no. 2, pp. 328--343, (1952).

\vskip 0.3cm

\noindent 12. L.L. Foldy, "Relativistic partivle systems with interaction," \textit{Phys. Rev.}, vol. 122, no. 1, pp. 275--288, (1961).

\vskip 0.3cm

\noindent 13. R.A. Weder, "Spectral properties of one-body relativistic spin-zero hamiltonians," \textit{Annal. de l'I. H.P., section A.}, vol. 20, no. 2, pp. 211--220, (1974).

\vskip 0.3cm

\noindent 14. R.A. Weder, "Spectral analysis of pseudodifferential operators*," \textit{J. Function. Analys.}, vol. 20, no. 4, pp. 319--337, (1975).

\vskip 0.3cm

\noindent 15. I.W. Herbst, "Spectral theory of the operator $(\textbf{p}^{2}+m^{2})^{1/2}-Z e^{2}/r$," \textit{Comm. Math. Phys.}, vol. 53, no. 3, pp. 285--294, (1977).

\vskip 0.3cm

\noindent 16. I.W. Herbst, "Addendum," \textit{Comm. Math. Phys.}, vol. 55, no. 3, p. 316, (1977).

\vskip 0.3cm

\noindent 17. P. Castorina, P. Cea, G. Nardulli, and G. Paiano, "Partial-wave analysis of a relativistic Coulomb problem," \textit{Phys. Rev. D.}, vol. 29, no. 11, pp. 2660--2662, (1984).

\vskip 0.3cm 

\noindent 18. L.J. Nickisch and L. Durand, "Salpeter equation in position space: Numerical solution for arbitrary confining potentials," \textit{Phys. Rev. D.}, vol. 30, no. 3, pp. 660--670, (1984).

\vskip 0.3cm 

\noindent 19. A. Martin and S.M. Roy, "Semi-relativistic stability and critical mass of a system of spinless bosons in gravitational interaction," \textit{Phys. Lett. B.}, vol. 223, no. 3,4, pp. 407--411, (1989).

\vskip 0.3cm

\noindent 20. J.C. Raynal, S.M. Roy, V. Singh, A. Martin and J. Stubble, "The <<Herbst Hamiltonian>> and the mass of boson stars," \textit{Phys. Lett. B.}, vol. 320, no. 1,2, pp. 105--109, (1994).

\vskip 0.3cm 

\noindent 21. L.P. Fulcher, "Matrix representation of the nonlocal kinetic energy operator, the spinless Salpeter equation and the Cornell potential," \textit{Phys. Rev. D.}, vol. 50, no. 1, pp. 447--453, (1994).

\vskip 0.3cm

\noindent 22. J.W. Norbury, K.M. Maung and D.E. Kahana, "Exact numerical solution of the spinless Salpeter equation for the Coulomb potential in momentum space," \textit{Phys. Rev. A.}, vol. 50, no. 5, pp. 3609--3613, (1994).

\vskip 0.3cm

\noindent 23. W. Lucha and F.F. Schobert, "Relativistic Coulomb problem: analytic upper bounds on energy levels," \textit{Phys. Rev. A.}, vol. 54, no. 5, pp. 3790--3794, (1996).

\vskip 0.3cm

\noindent 24. W. Lucha and F.F. Schobert, "Spinless Salpeter equation: Laguerre bounds on energy levels," \textit{Phys. Rev. A.}, vol. 56, no. 1, pp. 139--145, (1997).

\vskip 0.3cm

\noindent 25. F. Brau, "Integral equation formulation of the spinless Salpeter equation," \textit{J. Math. Phys.}, vol. 39, no. 4, pp. 2254--2263, (1998).

\vskip 0.3cm

\noindent 26. F. Brau, "Analytical solution of the relativistic Coulomb problem with a hard core interaction for a one-dimensional spinless Salpeter equation," \textit{J. Math. Phys.}, vol. 40, no. 3, pp. 1119--1126, (1999).

\vskip 0.3cm

\noindent 27. F. Brau, "Upper limit on the critical strength of central potentials in relativistic quantum mechanics," \textit{J. Math. Phys.}, vol. 46, no. 3, pp. 032305 (1--9), (2005).

\vskip 0.3cm

\noindent 28. F. Brau, "Lower bounds for the spinless Salpeter equation," \textit{J. Nonlin. Math. Phys.}, vol. 12, suppl. 1, pp. 86--96, (2005).

\vskip 0.3cm

\noindent 29. T.L. Gill and W.W. Zahary, "Analytic representation of the square-root operator," \textit{J. Phys. A.}, vol. 38, no. 11, pp. 2479--2496, (2005).

\vskip 0.3cm

\noindent 30. Y. Chargui, L. Chetouani and A. Trabelsi, "Analytical treatment of the one-dimensional Coulomb
problem for the spinless Salpeter equation," \textit{J. Phys. A.}, vol. 42, no. 35, pp. 355203 (1--10), (2009).

\vskip 0.3cm

\noindent 31. Y. Chargui, A. Trabelsi and L. Chetouani, "The one-dimensional spinless Salpeter Coulomb problem with minimal length," \textit{Phys. Lett. A.}, vol. 374, no. 22, pp. 2243--2247, (2010).

\vskip 0.3cm

\noindent 32. N.C. Petroni, "L$\mathrm{\acute{e}}$vy-Schr$\mathrm{\ddot{o}}$dinger wave packets," \textit{J. Phys. A.}, vol. 44, no. 16, pp. 165305 (1--27), (2011).

\vskip 0.3cm

\noindent 33. C. Semay, "An upper bound for asymmetrical spinless Salpeter equations," \textit{Phys. Lett. A.}, vol. 376, no. 33, pp. 2217--2221, (2012).

\vskip 0.3cm

\noindent 34. Y. Chargui and A. Trabelsi, "The zero-mass spinless Salpeter equation with a regularized inverse
square potential," \textit{Phys. Lett. A.}, vol. 377, no. 3-4, pp. 158--166, (2013).

\vskip 0.3cm

\noindent 35. P. Garbaczewski and V. Stephanovich, "L$\mathrm{\acute{e}}$vy flights and nonlocal quantum dynamics," \textit{J. Math. Phys.}, vol. 54, no. 7, pp. 072103 (1--34), (2013).

\vskip 0.3cm

\noindent 36. von J. Neumann, \textit{Mathematische Grundlagen der Quantenmechanik,} Berlin: Verlag von Julius Springer, 1932, 262 p.

\vskip 0.3cm

\noindent 37. P.A.M. Dirac, "Relativistic wave equations," \textit{Proc. R. Soc. Lond. A.}, vol. 155, no. 886, pp. 447--459, (1936)

\vskip 0.3cm

\noindent 38. H.J. Bhabha, "Relativistic wave equations for the elementary particles," \textit{Rev. Mod. Phys.}, vol. 17, no. 2-3, pp. 200--216, (1945).

\vskip 0.3cm

\noindent 39. D.L. Weaver, C.L. Hammer and R.H. Good Jr., "Description of a particle with arbitrary mass and spin," \textit{Phys. Rev.}, vol. 135, no. 1B, pp. B241--B248, (1964).

\vskip 0.3cm

\noindent 40. D.L. Pursey, "General theory of covariant particle equations," \textit{Ann. Phys. (N.Y.)}, vol. 32, no. 1, pp. 157--191, (1965).

\vskip 0.3cm

\noindent 41. P.M. Mathews, "Relativistic Schr$\mathrm{\ddot{o}}$dinger equations for particles of arbitrary spin," \textit{Phys. Rev.}, vol. 143, no. 4, pp. 978--985, (1966).

\vskip 0.3cm

\noindent 42. T.J. Nelson and R.H. Good Jr., "Second-quantization process for particles with any spin and with internal symmetry," \textit{Rev. Mod. Phys.}, vol. 40, no. 3, pp. 508--522, (1968).

\vskip 0.3cm

\noindent 43. R.F. Guertin, "Relativistic Hamiltonian equations for any spin," \textit{Ann. Phys. (N.Y.)}, vol. 88, no. 2, pp. 504--553, (1974).

\vskip 0.3cm

\noindent 44. R.A. Kraicik and M.M. Nieto, "Bhabha first-order wave equations. VI. Exact, closed-form,
Foldy-Wouthuysen transformations and solutions," \textit{Phys. Rev. D.}, vol. 15, no. 2, pp. 433--444, (1977).

\vskip 0.3cm

\noindent 45. R-K. Loide, I. Ots, and R. Saar, "Bhabha relativistic wave equations," \textit{J. Phys. A.}, vol. 30, no. 11, pp. 4005--4017, (1997).

\vskip 0.3cm

\noindent 46. D.M. Gitman and A.L. Shelepin, "Fields on the Poincar$\mathrm{\acute{e}}$ group: arbitrary spin
description and relativistic wave equations," \textit{Int. J. Theor, Phys.}, vol. 40, no. 3, pp. 603--684, (2001).

\vskip 0.3cm

\noindent 47. M. Fierz and W. Pauli, "On relativistic wave equations for particles of arbitrary spin in an electromagnetic field," \textit{Proc. R. Soc. Lond. A.}, vol. 173, no. 953, pp. 211--232, (1939).

\vskip 0.3cm

\noindent 48. W. Rarita and J. Schwinger, "On a theory of particles with half-integral spin," \textit{Phys. Rev.}, vol. 60, no. 1, p. 61, (1941).

\vskip 0.3cm

\noindent 49. A.S. Davydov, "Wave equations of a particle having spin 3/2 in the absence of a field," \textit{J. Exper. Theor. Phys.}, vol. 13, no. 9--10, pp. 313--319, (1943) (in Russian).

\vskip 0.3cm

\noindent 50. G. Velo and D. Zwanziger, "Propagation and quantization of Rarita-Schwinger waves
in an external electromagnetic potential," \textit{Phys. Rev.}, vol. 186, no. 5, pp. 1337--1341, (1969).

\vskip 0.3cm

\noindent 51. G. Velo and D. Zwanziger, "Noncausality and other defects of interaction Lagrangians for
particles with spin one and higher," \textit{Phys. Rev.).}, vol. 188, no. 5, pp. 2218--2222, (1969).

\vskip 0.3cm

\noindent 52. G. Velo and D. Zwanziger, "Fallacy of perturbative methods for higher-spin equations," \textit{Lett. Nuov. Cim.}, vol. 15, no. 2, pp. 39--40, (1976).

\vskip 0.3cm

\noindent 53. C. R. Hagen, "New inconsistencies in the quantization of spin-$\frac{3}{2}$ fields," \textit{Phys. Rev. D.}, vol. 4, no. 8, pp. 2204--2208, (1971).

\vskip 0.3cm

\noindent 54. J. D. Kimel and L. M. Nath, "Quantization of the spin-$\frac{3}{2}$ field in the presence of interactions. I," \textit{Phys. Rev. D.}, vol. 6, no. 8, pp. 2132--2144, (1972).

\vskip 0.3cm

\noindent 55. L. P. S. Singh, "Noncnusal propagation of classical Rarita-Schwinger waves," \textit{Phys. Rev. D.}, vol. 7, no. 4, pp. 1256--1258, (1973).

\vskip 0.3cm

\noindent 56. J. Prabhakaran, M. Seetharaman, and P. M. Mathews, "Causality of propagation of spin-$\frac{3}{2}$ fields coupled to spinor and scalar fields," \textit{Phys. Rev. D.}, vol. 12, no. 10, pp. 3191--3194, (1975).

\vskip 0.3cm

\noindent 57. M. Kobayashi* and A. Shamaly, "Minimal electromagnetic coupling for massive spin-two fields," \textit{Phys. Rev. D.}, vol. 17, no. 8, pp. 2179--2181, (1978).

\vskip 0.3cm

\noindent 58. A. Aurilia, M. Kobayashi and Y. Takahashi, "Remarks on the constraint structure and the quantization of the Rarita-Schwinger field," \textit{Phys. Rev. D.}, vol. 22, no. 6, pp. 1368--1374, (1980).

\vskip 0.3cm

\noindent 59. A.Z. Capri and R.L. Kobes, "Further problems in spin-3/2 field theories," \textit{Phys. Rev. D.}, vol. 22, no. 8, pp. 1967--1978, (1980).

\vskip 0.3cm

\noindent 60. C. Sierra, "Classical and quantum aspects of fields with secondary constraints," \textit{Phys. Rev. D.}, vol. 26, no. 10, pp. 2730--2744, (1982).

\vskip 0.3cm

\noindent 61. T. Darkhosh, "Is there a solution to the Rarita-Schwinger wave equation in the presence
of an external electromagnetic field?," \textit{Phys. Rev. D.}, vol. 32, no. 12, pp. 3251--3256, (1985).

\vskip 0.3cm

\noindent 62. V. Bargman and E.P. Wigner, "Group theoretical discussion of relativistic wave equations," \textit{Proc. Nat Acad. Sci. USA.}, vol. 34, no. 5, pp. 211--223, (1948)

\vskip 0.3cm

\noindent 63. D.L. Pursey, "A Foldy-Wouthuysen transformation for particles of arbitrary spin," \textit{Nucl. Phys.}, vol. 53, no. 1, pp. 174--176, (1964).

\vskip 0.3cm

\noindent 64. V.S. Vladimirov, \textit{Methods of the theory of generalized functions,} London: Taylor and Francis, 2002, 311 p.

\vskip 0.3cm

\noindent 65. N.N. Bogoliubov, A.A. Logunov and I.T. Todorov, \textit{Introduction to axiomatic quantum field theory,} Florida: W.A. Benjamin, Inc., 1975, 707 p.

\vskip 0.3cm

\noindent 66. P. Garbaczewski, "Boson-fermion duality in four dimensions: comments on the paper of Luther and Schotte," \textit{Int. J. Theor. Phys.}, vol. 25, no. 11, pp. 1193--1208, (1986).

\vskip 0.3cm

\noindent 67. W.I. Fushchich and A.G. Nikitin, \textit{Symmetries of equations of quantum mechanics,} New York: Allerton Press Inc., 1994, 480 p.

\vskip 0.3cm

\noindent 68. P. Lounesto, \textit{Clifford algebras and spinors, 2-nd edition} Cambridge: Cambridge University Press, 2001, 338 p.

\vskip 0.3cm
	
\noindent 69. V.M. Simulik and I.Yu. Krivsky, "On the extended real Clifford--Dirac algebra and new physically
meaningful symmetries of the Dirac equation with nonzero mass," \textit{Reports of the National Academy of Sciences of Ukraine}, no. 5, pp. 82--88, (2010) (in Ukrainian).

\vskip 0.3cm

\noindent 70. V.M. Simulik and I.Yu. Krivsky, "Bosonic symmetries of the Dirac equation," \textit{Phys. Lett. A.}, vol. 375, no. 25, pp. 2479--2483, (2011).

\vskip 0.3cm

\noindent 71. V.M. Simulik, I.Yu. Krivsky,  and I.L. Lamer, "Some statistical aspects of the spinor field
Fermi--Bose duality," \textit{Cond. Matt. Phys.}, vol. 15, no. 4, pp. 43101 (1--10), (2012).

\vskip 0.3cm
 
\noindent 72. V.M. Simulik, I.Yu. Krivsky,  and I.L. Lamer, "Application of the generalized Clifford--Dirac algebra to the proof of the Dirac equation Fermi--Bose duality," \textit{TWMS J. App. Eng. Math.}, vol. 3, no. 1, pp. 46--61, (2013).

\vskip 0.3cm

\noindent 73. V.M. Simulik, I.Yu. Krivsky,  and I.L. Lamer, "Bosonic symmetries, solutions and conservation laws for the Dirac equation with nonzero mass," \textit{Ukr. J. Phys.}, vol. 58, no. 6, pp. 523--533, (2013).

\vskip 0.3cm

\noindent 74. J. Elliott and P. Dawber, \textit{Symmetry in Physics, vol.1,} London: Macmillian Press, 1979, 366 p.

\vskip 0.3cm

\noindent 75. B. Wybourne,, \textit{Classical groups for Physicists} New York: John Wiley and sons, 1974, 415 p.

\vskip 0.3cm

\noindent 76. W. Pauli, "On the conservation of the lepton charge," \textit{Nuov. Cim.}, vol. 6, no. 1, pp. 204--215, (1957).

\vskip 0.3cm

\noindent 77. F. G$\ddot{\mathrm{u}}$rsey, "Relation of charge independence and baryon conservation to Pauli's transformation," \textit{Nuov. Cim.}, vol. 7, no. 3, pp. 411--415, (1958).

\vskip 0.3cm

\noindent 78. N. Kh. Ibragimov, "Invariant variational problems and conservation laws (remarks on Noether's theorem)," \textit{Theor. Math. Phys.}, vol. 1, no. 3, pp. 267--274, (1969).

\vskip 0.3cm

\noindent 79. W.A. Hepner, "The inhomogeneous Lorentz group and the conformal group," \textit{Nuov. Cim.}, vol. 26, no. 2, pp. 351--368, (1962).

\vskip 0.3cm

\noindent 80. M. Petras, "The SO(3,3) group as a common basis for Dirac's and Proca's equations," \textit{Czech. J. Phys.}, vol. 46, no. 6, pp. 455--464, (1995).

\vskip 0.3cm

\noindent 81. I. Yu. Krivsky and V.M. Simulik, "The Dirac equation and spin 1 representations, a connection with symmetries of the Maxwell equations," \textit{Theor. Math. Phys.}, vol. 90, no. 3, pp. 265--276, (1992).

\vskip 0.3cm

\noindent 82. I. Yu. Krivsky, R.R. Lompay, and V.M. Simulik, "Symmetries of the complex Dirac--K$\ddot{\mathrm{a}}$hler equation," \textit{Theor. Math. Phys.}, vol. 143, no. 1, pp. 541--558, (2005).

\vskip 0.3cm

\noindent 83. D.F. Styer et al., "Nine formulations of quantum mechanics," \textit{Am. J. Phys.}, vol. 70, no. 3, pp. 288--297, (2002).

\vskip 0.3cm

\noindent 84. B. Thaller, \textit{The Dirac equation} Berlin: Springer, 1992, 358 p.

\vskip 0.3cm

\noindent 85. N.N. Bogoliubov and D.V. Shirkov, \textit{Introduction to the theory of quantized fields,} New York: John Wiley and Sons, 1980, 620 p.

\vskip 0.3cm

\noindent 86. S.I. Kruglov, "Dirac--K$\ddot{\mathrm{a}}$hler equation," \textit{Int. J. Theor. Phys.}, vol. 41, no. 4, pp. 653--687, (2002).

\vskip 0.3cm

\noindent 87. V.P. Neznamov and A.J. Silenko, "Foldy--Wouthuysen wave functions and conditions of transformation between Dirac and Foldy--Wouthuysen representations," \textit{J. Math. Phys.}, vol. 50, no. 12, pp. 122302(1--15), (2009).

\vskip 0.3cm

\noindent 88. J. Sakurai, \textit{Advanced quantum mechanics} New York: Addison--Wesley Pub. Co, 1967, 335 p.

\vskip 0.3cm

\noindent 89. L. Ryder, \textit{Quantum field theory, 2-nd edit.} Cambridge: University Press, 1996, 487 p.

\vskip 0.3cm

\noindent 90. J. Keller, \textit{Theory of the electron. A theory of matter from START} Dordrecht: Kluwer Academic Publishers, 2001, 257 p.

\vskip 0.3cm

\noindent 91. V. Fock and D. Iwanenko, "Uber eine mogliche geometrische Deutung der relativistischen Quantentheorie," \textit{Z. Phys.}, vol. 54, no. 11-12, pp. 798--802, (1929).

\vskip 0.3cm

\noindent 92. V. Fock, "Geometrsierung der Diracschen Theorie des Elektrons," \textit{Z. Phys.}, vol. 57, no. 3-4, pp. 261--277, (1929).

\vskip 0.3cm

\noindent 93. A. Wightman, \textit{in <<Dispersion relations and elementary particles>>, edit. C. De Witt and R. Omnes.} New York: Wiley and Sons, 1960, 671 p.

\vskip 0.3cm

\noindent 94. H. Sallhofer, "Elementary derivation of the Dirac equation. I," \textit{Z. Naturforsch. A.}, vol. 33, no. 11, pp. 1378--1380, (1978).

\vskip 0.3cm

\noindent 95. H. Sallhofer, "Elementary derivation of the Dirac equation. X," \textit{Z. Naturforsch. A.}, vol. 41, no. 3, pp. 468--470, (1986).

\vskip 0.3cm

\noindent 96. S. Strinivasan and E. Sudarshan, "A direct derivation of the Dirac equation via quaternion measures," \textit{J. Phys. A.}, vol. 29, no. 16, pp. 5181--5186, (1996).

\vskip 0.3cm

\noindent 97. L. Lerner, "Derivation of the Dirac equation from a relativistic representation of spin," \textit{Eur. J. Phys.}, vol. 17, no. 4, pp. 172--175, (1996).

\vskip 0.3cm

\noindent 98. T. Kubo, I. Ohba, and H. Nitta, "A derivation of the Dirac equation in an external field based on the Poisson process," \textit{Phys. Lett. A.}, vol. 286, no. 4, pp. 227--230, (2001).

\vskip 0.3cm

\noindent 99. H. Cui, "A method for deriving the Dirac equation from the relativistic Newton's second law," arXiv: 0102114v4 [quant-ph] 15 Aug 2001, 4 p.

\vskip 0.3cm

\noindent 100. Y.J. Ng and H. van Dam, "A geometrical derivation of the Dirac equation," arXiv: 0211002v2 [hep-th] 4 Feb 2003, 9 p.

\vskip 0.3cm

\noindent 101. M. Calerier and L. Nottale, "A scale relativistic derivation of the Dirac equation," \textit{Electromagnetic Phenomena}, vol. 3, no. 1(9), pp. 70--80, (2003).

\vskip 0.3cm

\noindent 102. M. Evans, "A generally covariant field equation for gravitation and electromagnetism," \textit{Found. Phys. Lett.}, vol. 16, no. 4, pp. 369--377, (2003).

\vskip 0.3cm

\noindent 103. M. Evans, "Derivation of Dirac's equation from the Evans wave equation" \textit{Found. Phys. Lett.}, vol. 17, no. 2, pp. 149--166, (2004).

\vskip 0.3cm

\noindent 104. V.M. Simulik and I.Yu. Krivsky, "Relationship between the Maxwell and Dirac equations: symmetries, quantization, models of atom," \textit{Rep. Math. Phys.}, vol. 50, no. 3, pp. 315--328, (2002).

\vskip 0.3cm

\noindent 105. V.M. Simulik and I.Yu. Krivsky, "Classical electrodynamical aspect of the Dirac equation," \textit{Electromagnetic Phenomena},  vol. 3, no. 1(9), pp. 103--114, (2003).

\vskip 0.3cm
 
\noindent 106. V.M. Simulik, \textit{in <<What is the electron?>>, edit. V. Simulik.} Montreal: Apeiron, 2005, 282 p.

\vskip 0.3cm

\noindent 107. F. Gaioli and E. Alvarez, "Some remarks about intrinsic parity in Ryder's derivation of the Dirac equation" \textit{Am. J. Phys.},  vol. 63, no. 2, pp. 177--178, (1995).

\vskip 0.3cm

\noindent 108. S. Efthimiades, "Physical meaning and derivation of Schrodinger and Dirac equations," arXiv: 0607001v3 [quant-ph] 4 Jul 2006, 8 p.

\vskip 0.3cm

\noindent 109. G.M. D'Ariano and P. Perinotti, "Derivation of the Dirac equation from principles of information processing," arXiv: 1306.1934v2 [quant-ph] 28 Dec 2014, 20 p.

\vskip 0.3cm

\noindent 110. H. Bondi, "Negative mass in general relativity," \textit{Rev. Mod. Phys.}, vol. 29, no. 3, pp. 423--428, (1957).

\vskip 0.3cm

\noindent 111. E. Recami and G. Zino, "About new space-time symmetries in relativity and quantum mechanics," \textit{Nuovo Cim. A.}, vol. 33, no. 2, pp. 205--215, (1976).

\vskip 0.3cm

\noindent 112. G. Landis, "Comments on negative mass propulsion," \textit{J. Propulsion and Power}, vol. 7, no. 2, p. 304, (1990).

\vskip 0.3cm

\noindent 113. R. Wayne, "Symmetry and the order of events in time, a proposed identity of negative mass with antimatter," \textit{Turk. J. Phys.}, vol. 36, no. 2, pp. 165--177, (2012).

\vskip 0.3cm

\noindent 114. W. Kuellin et al, "Coherent ballistic motion of electrons in a periodic potential," \textit{Phys. Rev. Lett.}, vol. 104, no. 14, pp. 146602(1-4), (2010).

\vskip 0.3cm

\noindent 115. V.P. Neznamov, "On the theory of interacting fields in the Foldy--Wouthuysen representation," \textit{Phys. Part. Nucl.}, vol. 37, no. 1, pp. 86--103, (2006).

\vskip 0.3cm

\noindent 116. A.J. Silenko, "Scalar particle in general inertial and gravitational fields and conformal invariance revisited," \textit{Phys. Rev. D.}, vol. 88, no. 4, pp. 045004(1--5), (2013).

\vskip 0.3cm

\noindent 117. A.J. Silenko, "Foldy--Wouthyusen transformation and semiclassical limit for relativistic particles in strong external fields," \textit{Phys. Rev. A.}, vol. 77, no. 1, pp. 012116(1--7), (2008).

\vskip 0.3cm

\noindent 118. A.J. Silenko, "Quantum-mechanical description of spin-1 particles with electric dipole moments," \textit{Phys. Rev. D.}, vol. 87, no. 7, pp. 073015(1--5), (2013).

\vskip 0.3cm

\noindent 119. A.J. Silenko, "High precision description and new properties of a spin-1 particle in a magnetic field," \textit{Phys. Rev. D.}, vol. 89, no. 12, pp. 121701(R)(1--6), (2014).

\vskip 0.3cm

\noindent 120. K.M. Case, "Some generalizations of the Foldy--Wouthuysen transformation," \textit{Phys. Rev.}, vol. 95, no. 5, pp. 1323--1328, (1954).

\vskip 0.3cm

\noindent 121. P. Caban, J. Rembielinski and M. Wlodarczyk,, "Spin operator in the Dirac theory," \textit{Phys. Rev. A.)}, vol. 88, no. 2, pp. 022119(1--8), (2013).

\vskip 0.3cm

\noindent 122. I.Yu. Krivsky, V.M. Simulik, T.M. Zajac and I.L. Lamer, Derivation of the Dirac and Maxwell equations from the first principles of relativistic canonical quantum mechanics, Proc. of the 14-th Internat. Conference <<Mathematical Methods in Electromagnetic Theory>>, 28--30 August 2012, Institute of Radiophysics and Electronics, Kharkiv, Ukraine. P. 201--204.

\vskip 0.3cm

\noindent 123. V.M. Simulik, Electromagnetic field equations, which follow from the relativistic canonical quantum mechanics, Proc. of the 15-th Internat. Conference <<Mathematical Methods in Electromagnetic Theory>>, 26--28 August 2014, O. Honchar National University, Dnipropetrovsk, Ukraine. P. 9--30.

\vskip 0.3cm

\noindent 124. V.M. Simulik, Link between the relativistic canonical quantum mechanics of arbitrary spin
and the corresponding feld theory, Book of Abstracts of the XXIIIth International Conference on Integrable Systems and Quantum symmetries, 23--27 June 2015, Prague, Czech Republic, P. 110.

\vskip 0.3cm

\noindent 125. V.M. Simulik, Relativistic description of the spin S=3/2 particle-antiparticle doublet interacting with electromagnetic field, Abstracts of the 8-th Symposium on Integrable Systems, 3--4 July 2015, Lodz, Poland, P. 23.

\end{document}